\documentclass[prd,aps,twocolumn,a4paper,floatfix,showpacs,nofootinbib,superscriptaddress,nolongbibliography]{revtex4-2}

\usepackage[utf8]{inputenc}
\usepackage[T1]{fontenc}
\usepackage{graphicx,psfrag}
\usepackage{mathrsfs}
\usepackage{amsmath,amsfonts,amssymb,pifont,gensymb}
\usepackage{multirow,enumerate}
\usepackage{comment,hyperref}
\usepackage{color}
\usepackage{acronym}
\usepackage{xspace}
\usepackage[normalem]{ulem}
\usepackage{mathtools}
\usepackage{subfigure}
\usepackage{makecell}
\usepackage{braket}

\newacro{BH}{black hole}
\newacro{NS}{neutron star}
\newacro{PN}{Post-Newtonian}
\newacro{BBH}{binary black hole}
\newacro{BNS}{binary neutron star}
\newacro{EOB}{effective-one-body}
\newacro{NR}{numerical relativity}
\newacro{GW}{gravitational wave}
\newacro{EOS}{equation-of-state}

\newcommand{\be}{\begin{equation}}
\newcommand{\ee}{\end{equation}}
\newcommand{\bea}{\begin{eqnarray}}
\newcommand{\eea}{\end{eqnarray}}
\newcommand{\bel}{\begin{align}}
\newcommand{\eel}{\end{align}}

\newcommand{\Msun}{M_\odot}

\newcommand{\phenxas}{\textmd{IMRPhenomXAS}\xspace}
\newcommand{\phenxp}{\textmd{IMRPhenomXP}\xspace}

\newcommand{\phenxastidal}{\textmd{IMRPhenomXAS\_NRTidalv2}\xspace}
\newcommand{\phenxptidal}{\textmd{IMRPhenomXP\_NRTidalv2}\xspace}
\newcommand{\phendtidal}{\textmd{IMRPhenomD\_NRTidalv2}\xspace}
\newcommand{\phenptidal}{\textmd{IMRPhenomPv2\_NRTidalv2}\xspace}

\newcommand{\teob}{\textmd{TEOBResumS-GIOTTO}\xspace}

\def\i{{\rm i}}

\def\Msun{{\rm M_{\odot}}}
\def\GMc2{{\rm G M_{\odot} c^{-2}}}

\def\SEOBNRv4T{\texttt{SEOBNRv4T}\xspace}

\newcommand{\UIB}{Departament de F\'isica, Universitat de les Illes Balears, IAC3 -- IEEC, Crta. Valldemossa km 7.5, E-07122 Palma, Spain}
\newcommand{\UP}{Institut f\"{u}r Physik und Astronomie, Universit\"{a}t Potsdam, Haus 28, Karl-Liebknecht-Str. 24/25, 14476, Potsdam, Germany}
\newcommand{\AEI}{Max Planck Institute for Gravitational Physics (Albert Einstein Institute), Am M\"uhlenberg 1, Potsdam 14476, Germany}
\newcommand{\UM}{Department of Physics and Astronomy, University of Mississippi, University, Mississippi 38677, USA}
\newcommand{\bham}{School of Physics and Astronomy, University of Birmingham, Edgbaston, Birmingham, B15 2TT, United Kingdom}
\newcommand{\bhamgw}{Institute for Gravitational Wave Astronomy, University of Birmingham, Edgbaston, Birmingham, B15 2TT, United Kingdom}
\newcommand{\nikh}{Nikhef, Science Park 105, 1098 XG Amsterdam, The Netherlands}

\begin{document}

\title{New gravitational waveform
    model for precessing binary neutron stars with double-spin effects}

\author{Marta Colleoni}
\affiliation{\UIB}

\author{Felip A. Ramis Vidal}
\affiliation{\UIB}

\author{Nathan~K.~\surname{Johnson-McDaniel}}
\affiliation{\UM}

\author{Tim Dietrich}
\affiliation{\UP}
\affiliation{\AEI}

\author{Maria Haney}
\affiliation{\nikh}

\author{Geraint Pratten}
\affiliation{\bham}
\affiliation{\bhamgw}

\date{\today}

\begin{abstract}
We present two new frequency-domain gravitational waveform models for the analysis of signals emitted by binary neutron star coalescences: \textmd{IMRPhenomXAS\_NRTidalv2} and \textmd{IMRPhenomXP\_NRTidalv2}. Both models are available through the public algorithm library \textmd{LALSuite} and represent the first extensions of \textmd{IMRPhenomX} models including matter effects.  We show here that these two models represent a significant advancement in efficiency and accuracy with respect to their phenomenological predecessors, \textmd{IMRPhenomD\_NRTidalv2} and \textmd{IMRPhenomPv2\_NRTidalv2}. The computational efficiency of the new models is achieved through the application of the same multibanding technique previously applied to binary black hole models. Furthermore, \textmd{IMRPhenomXP\_NRTidalv2} implements a more accurate description of the precession dynamics, including double-spin effects and, optionally, matter effects in the twisting-up construction. The latter are available through an option to use a numerical integration of the post-Newtonian precession equations. We show that the new precession descriptions allow the model to better reproduce the phenomenology observed in numerical-relativity simulations of precessing binary neutron stars. Finally, we present some applications of the new models to Bayesian parameter estimation studies, including a reanalysis of GW170817 and a study of simulated observations using numerical relativity waveforms for nonprecessing binary neutron stars with highly spinning components. We find that in these cases the new models make a negligible difference in the results. Nevertheless, by virtue of the aforementioned improvements, the new models represent valuable tools for the study of future detections of coalescing binary neutron stars.
\end{abstract}

\maketitle

\section{Introduction}
\label{sec:intro}

The observation of gravitational waves (GWs) from the binary neutron star merger (BNS) GW170817~\cite{LIGOScientific:2017vwq} together with its electromagnetic counterparts, the kilonova AT2017gfo, the short gamma-ray burst GRB170817A and its afterglow, have been a breakthrough in the field of multi-messenger astronomy~\cite{LIGOScientific:2017zic,LIGOScientific:2017ync}. One requires accurate theoretical models for the GWs and electromagnetic emission from such binaries in order to extract information about their properties from the multimessenger observations. In particular, such observations allow us to constrain the properties of neutron-rich material at greater than nuclear densities, i.e., the equation of state (EOS) governing the neutron star interior; see, e.g.,~\cite{LIGOScientific:2017vwq,LIGOScientific:2018hze,LIGOScientific:2018cki,De:2018uhw,LIGOScientific:2019eut,Capano:2019eae,Essick:2019ldf,Narikawa:2019xng,Dietrich:2020efo,Landry:2020vaw}.

Constraints on the EOS from GW observations are derived from measurements of the tidal effects in the last stages of the coalescence~\cite{Flanagan:2007ix,Damour:2009wj,Damour:2012yf,Henry:2020ski}, i.e., the tidal deformations of the individual stars caused by the presence of their companion~\cite{Hinderer:2007mb,Hinderer:2009ca,Damour:2009vw,Binnington:2009bb}. These deformations, characterized by the dimensionless tidal deformability $\Lambda$, modify the inspiral dynamics and the emitted GW signal. Most notably, the inspiral is accelerated due to the additional attractive tidal interaction. This leads to an earlier merger of the stars compared to a coalescence of binary black holes; see~\cite{Dietrich:2020eud,Chatziioannou:2020pqz} for recent reviews, including many references, and~\cite{Chatziioannou:2021tdi,Ghosh:2022muc,Jiang:2022tps} for a selection of more recent work. \\

In this article, we present two new phenomenological models for the GWs from BNSs to improve our ability to interpret BNS systems through GW astronomy. In particular, we focus on improving the description of BNS systems in which the individual stars also have an intrinsic rotation, and the rotation axis is not aligned with the orbital angular momentum. In such systems, the orbital plane will start precessing, which leads to a more complex GW spectrum than for aligned-spin cases. One expects that there will be some spin misalignment in BNS systems formed through the isolated binary channel, due to the kicks imparted on the neutron stars by the supernova that formed them, and thus a measurement of this spin misalignment can provide constraints on neutron star natal kicks; see, e.g., \cite{Postnov:2007yv}. One expects more significant spin misalignments in the dynamical formation channel, but this is expected to contribute negligibly to the rate of BNS mergers (see~\cite{Ye:2019xvf}).

The two models we present here, \textmd{IMRPhenomXAS\_NRTidalv2} and \textmd{IMRPhenomXP\_NRTidalv2}, are based on the \textmd{IMRPhenomX} framework for binary black hole waveforms~\cite{Pratten:2020fqn,Garcia-Quiros:2020qlt,Pratten:2020ceb} and on the NRTidalv2 tidal extension~\cite{Dietrich:2017aum, Dietrich:2018uni, Dietrich:2019kaq}. They are available through the \textsc{LALSuite} package~\cite{lalsuite} and are ready for use in GW data analysis. 
Our new models naturally inherit all the properties of the IMRPhenomX waveform family, as far as the binary black hole baseline is concerned. In what follows, we briefly discuss some of the main differences between the models presented here and previous NRTidal implementations based on IMRPhenomD and IMRPhenomPv2. The first difference is in the aligned-spin binary black hole model to which tidal corrections are added. The models presented here rely on IMRPhenomXAS~\cite{Pratten:2020fqn}, which represents a substantial update to IMRPhenomD~\cite{Khan:2015jqa} in terms of robustness and accuracy. In phenomenological models, the extension to precessing systems is performed following the twisting-up construction first implemented in IMRPhenomPv2~\cite{Schmidt:2010it,Schmidt:2012rh,Hannam:2013oca}, where precessing waveforms are decomposed into a co-precessing frame signal, which is approximated by an aligned-spin waveform, and a set of Euler angles specifying the mapping between co-precessing and inertial frame. In IMRPhenomPv2, closed-form expressions for the Euler angles are computed using a single-spin approximation of the next-to-next-to-leading order post-Newtonian spin-precession equations, together with the stationary phase approximation~\cite{Bohe:PPv2}. The more recent IMRPhenomXP model~\cite{Pratten:2020ceb} builds upon the framework laid by IMRPhenomPv2, but significantly improves upon it by capturing double-spin effects in the precession dynamics. By default, the Euler angles used in the twisting up are computed following an approximate analytic treatment (precession averaged plus the leading correction) of the second post-Newtonian (2PN) precession dynamics, which offers an excellent trade-off between accuracy and computational efficiency for low-mass systems. Optionally, the user can activate a more accurate twisting-up prescription based on the numerical solution of the orbit-averaged \textmd{SpinTaylorT4} post-Newtonian equations~\cite{SpinTaylor_TechNote} as implemented in \textsc{LALSuite}, with 3.5PN point-particle phasing~\cite{Buonanno:2009zt}, 3PN spin effects in the phasing and precession dynamics~\cite{Bohe:2013cla,Bohe:2015ana}, and 6PN tidal effects in the phasing~\cite{Vines:2011ud} employed by default. Besides being more accurate than the approximate analytic description (both removing the approximation used to solve the precession equations and including higher-order PN terms), this option accounts for the presence of matter, notably in the treatment of spin-induced quadrupole moments in the precession dynamics, which are instead specialized to the black-hole case in the precession-averaged treatment. We note that, thanks to the modularity of \textmd{PhenomX}, any precessing option available for the binary-black hole model can be applied to its tidal extensions. Further details regarding the application of the ``SpinTaylor'' option to the binary black hole case are given in a companion paper~\cite{Colleoni:2024knd}. The wider spectrum of physical effects modeled by \textmd{IMRPhenomXP\_NRTidalv2} does not compromise its numerical efficiency. This is due to the application of the multibanding technique \cite{Garcia-Quiros:2020qlt}, which makes the model's evaluation time sit consistently well below that of other waveform models for BNS in the aligned-spin or precession-averaged cases.

This paper is structured as follows. In Sec.~\ref{sec:model}, we briefly describe how the models are constructed.
In Sec.~\ref{sec:tests}, we compare our new models to other approximants including matter effects, through benchmarks and match studies; we also show some visual comparisons to numerical relativity (NR) simulations of precessing BNSs. In Sec.~\ref{sec:pe}, we discuss some concrete applications of our models to parameter estimation (PE) studies of real and simulated GW signals. We present our conclusions in Sec.~\ref{sec:conclusions}. Finally, we describe the different settings available for the model in Appendix~\ref{app:settings} and give additional illustrations of the effects of settings on the accuracy of SpinTaylor Euler angles in Appendix~\ref{app:coarse_angles}. Throughout the paper, we employ geometric units where $G=c=1$.

\section{Model Description}
\label{sec:model}

\subsection{NRTidalv2: overview and implementation details}

The NRTidalv2 model provides closed-form expressions capturing matter effects that can be added to any frequency-domain binary black hole approximant. The amplitude of NRTidalv2 builds upon the model proposed in Kawaguchi~\emph{et al.}~\cite{Kawaguchi:2018gvj}, introducing an extra NR-calibrated term with the purpose of modeling unknown higher-order PN terms. The high-frequency part of the signal is suppressed by means of a Planck taper, starting from the estimated merger frequency, $f_{\rm{merg}}$, and ending at $1.2\,f_{\rm{merg}}$. The tidal phase is modeled with a Padé approximant, inspired by resummation techniques employed by effective-one-body (EOB) models~\cite{Bini:2012gu, Damour:2012yf}, where some of the coefficients are constrained thanks to analytical PN information, while others are informed by non-spinning, equal-mass NR simulations (see Sec.~II B of~\cite{Dietrich:2019kaq} for further details) and EOB waveforms~\cite{Nagar:2018zoe}. The phase model also includes quadratic and cubic-in-spin analytical corrections due to the spin-induced quadrupole and octupole deformations, first entering at 2PN and 3.5PN order, respectively. The coefficients parametrizing the spin-induced multipole moments are expressed in terms of the components' dimensionless tidal deformabilities through quasi-universal EOS independent relations \cite{Yagi:2016bkt}.

We highlight here some subtle differences between the NRTidal implementations for IMRPhenomD/Pv2 and IMRPhenomX. The IMRPhenomX models only support the NRTidalv2 extension and not the original NRTidal one.\footnote{In other words, the user will not be able to call IMRPhenomXAS\_NRTidal or IMRPhenomXP\_NRTidal, while the original NRTidal extension is still supported by IMRPhenomD and IMRPhenomPv2.}
Furthermore, in the NRTidal[v2] implementations based on IMRPhenomD, the spin-induced quadrupole terms were included as part of the PN coefficients used in the inspiral regime, except for the $3.5$PN terms in NRTidalv2, which were added for all frequencies. In the IMRPhenomX models, on the other hand, the spin-induced quadrupole terms are added over the model's entire frequency range. Since the transition to the intermediate frequency range of Phenom models (either IMRPhenomD or IMRPhenomX) is at quite high frequencies for BNS signals (at $\sim 1300$~Hz for IMRPhenomD and a binary with a total mass of $2.8\,\Msun$), this will likely not make a significant difference for current detector sensitivities. Finally, in the PhenomX NRTidalv2 implementation, both aligned-spin and precessing models include the tidal corrections to the amplitude of the (2,2) mode, whereas previously these were only included in \phenptidal.

\subsection{Dependence of Euler angles on the EOS}
\label{subsec:euler_angles}

\begin{figure}[h]
\begin{center}
\includegraphics[width=\columnwidth]{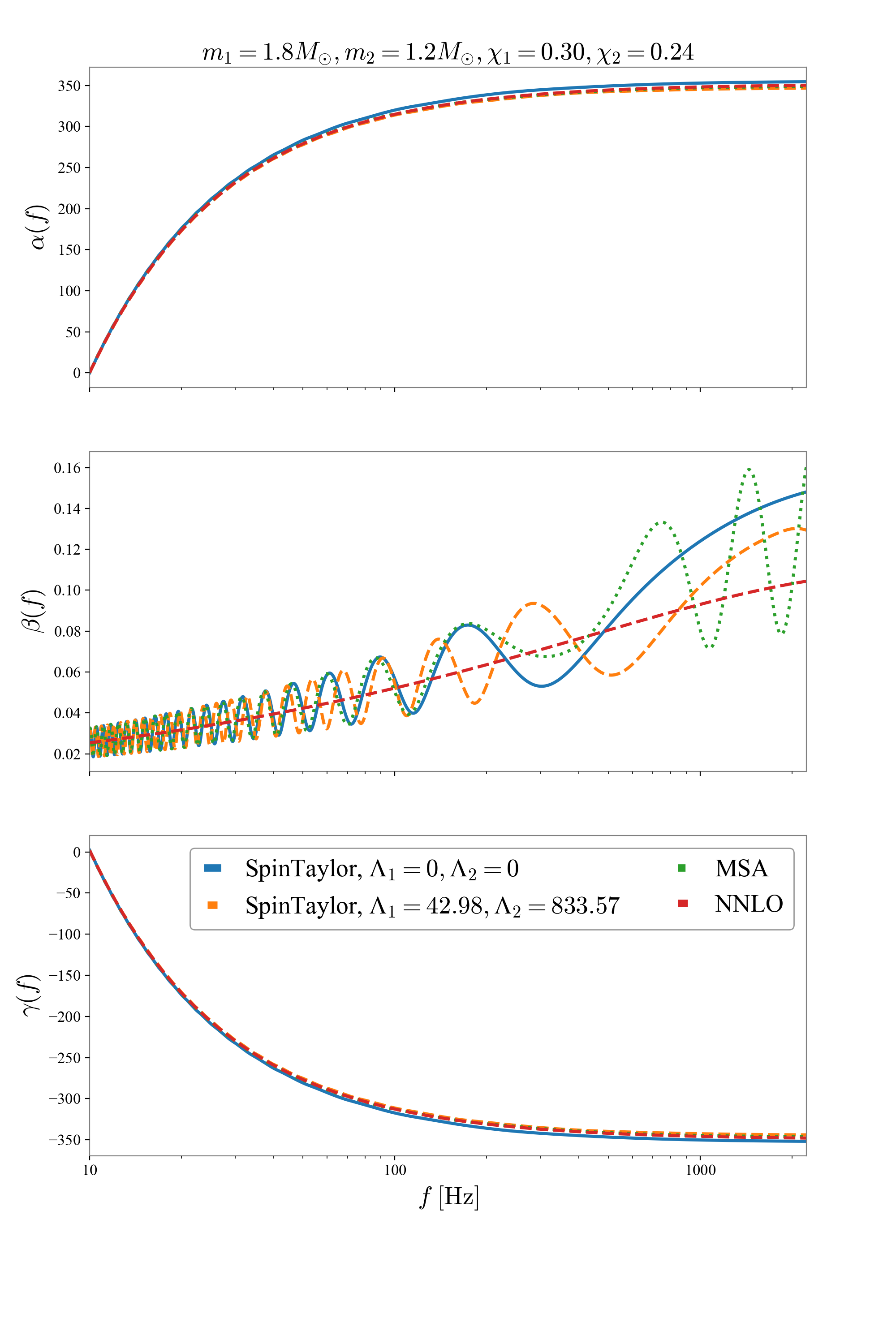}
\caption{The Euler angles $\alpha$, $\beta$, and $\gamma$ as a function of GW frequency, over the inspiral range of a BNS with masses $m_1=1.8\,\Msun$, $m_2=1.2\,\Msun$ and spins $\chi_1=(0.1,0.2,0.2)$, $\chi_2=(0.1,0.1,-0.2)$ at 10 Hz. The solid blue and orange dashed lines represent SpinTaylor solutions, where in the first case we set both tidal deformabilities to zero, while in the second we use the tidal deformabilities $\Lambda_A$ given by the SLy EOS~\cite{Douchin:2001sv} for these masses. The green dotted and red dash-dotted lines represent the MSA and NNLO solutions respectively. \label{fig:euler_angles}
}
\end{center}
\end{figure}

By default, the Euler angles employed in the twisting-up construction of IMRPhenomXP\_NRTidalv2 are computed following the method outlined in Chatziioannou~\emph{et al.}~\cite{Chatziioannou:2017tdw}, where the inspiral dynamics is solved within the post-Newtonian framework with the aid of multiple scale analysis (MSA). Here solutions to the precession-averaged equations and the leading correction to the precession averaging are efficiently evolved over the entire inspiral exploiting the separation between the radiation-reaction and the precession timescales. 
The analytical solution of \cite{Chatziioannou:2017tdw} is obtained under the hypothesis of quasicircularity and it is insensitive to matter effects, since it specializes to black hole spin-induced quadrupolar deformations, neglecting the EOS-dependent differences in the deformations present for neutron stars. The SpinTaylor option of IMRPhenomXP\_NRTidalv2 allows us to incorporate such effects into the twisting-up construction, albeit at the price of a higher computational cost (see Sec.~\ref{subsec:benchmarks}). We demonstrate these effects in Fig.~\ref{fig:euler_angles}, where we show the Euler angles computed via the MSA and SpinTaylor precession prescriptions for a BNS with mass ratio $1/q=1.5$, total mass $3\,\Msun$, and moderate spins on both components ($\chi_1=0.30$, $\chi_2=0.24$). For the SpinTaylor option, we show two different solutions: the first, which is represented by a blue solid line in Fig.~\ref{fig:euler_angles}, was computed setting both tidal deformabilities to zero (and thus the spin-induced quadrupoles to their black hole values), while the second (dashed orange line) corresponds to a BNS whose tidal deformabilities are those obtained from the binary's masses using the SLy EOS~\cite{Douchin:2001sv}, computed using LALSimulation~\cite{lalsuite}. In the same figure, we also show the MSA and next-to-next-to-leading order (NNLO) Euler angles: the latter corresponds to the solutions implemented in IMRPhenomPv2\_NRTidalv2.
 
 The two top panels show $\alpha$ and $\beta$, which measure the azimuthal and polar angles of the Newtonian orbital angular momentum in the $\hat{\textbf{J}}$-frame, where the $\hat{\textbf{z}}$ axis is aligned with the total angular momentum (see, e.g., Fig.~1 of \cite{Bohe:PPv2}), while the bottom one shows $\gamma$, which completes the triad of Euler angles specifying the mapping between the  co-precessing and $\hat{\textbf{J}}$ frame. The MSA and SpinTaylor solutions are in good agreement at low frequencies and, unlike the NNLO solution, capture the modulations in the opening angle of the precession cone $\beta$ due to double-spin effects (see middle panel). There is a progressive dephasing between different double-spin solutions as the GW frequency increases, which is enhanced when accounting for the tidal deformabilities and spin-induced quadrupoles of the two objects. In Sec.~\ref{subsec:precessing_bns}, we will investigate how these differences affect the full GW signal.

\subsection{Coarse grid integration}
\label{subsec:coarse_grid_integration}

\begin{figure}[h]
\begin{center}
\includegraphics[width=\columnwidth]{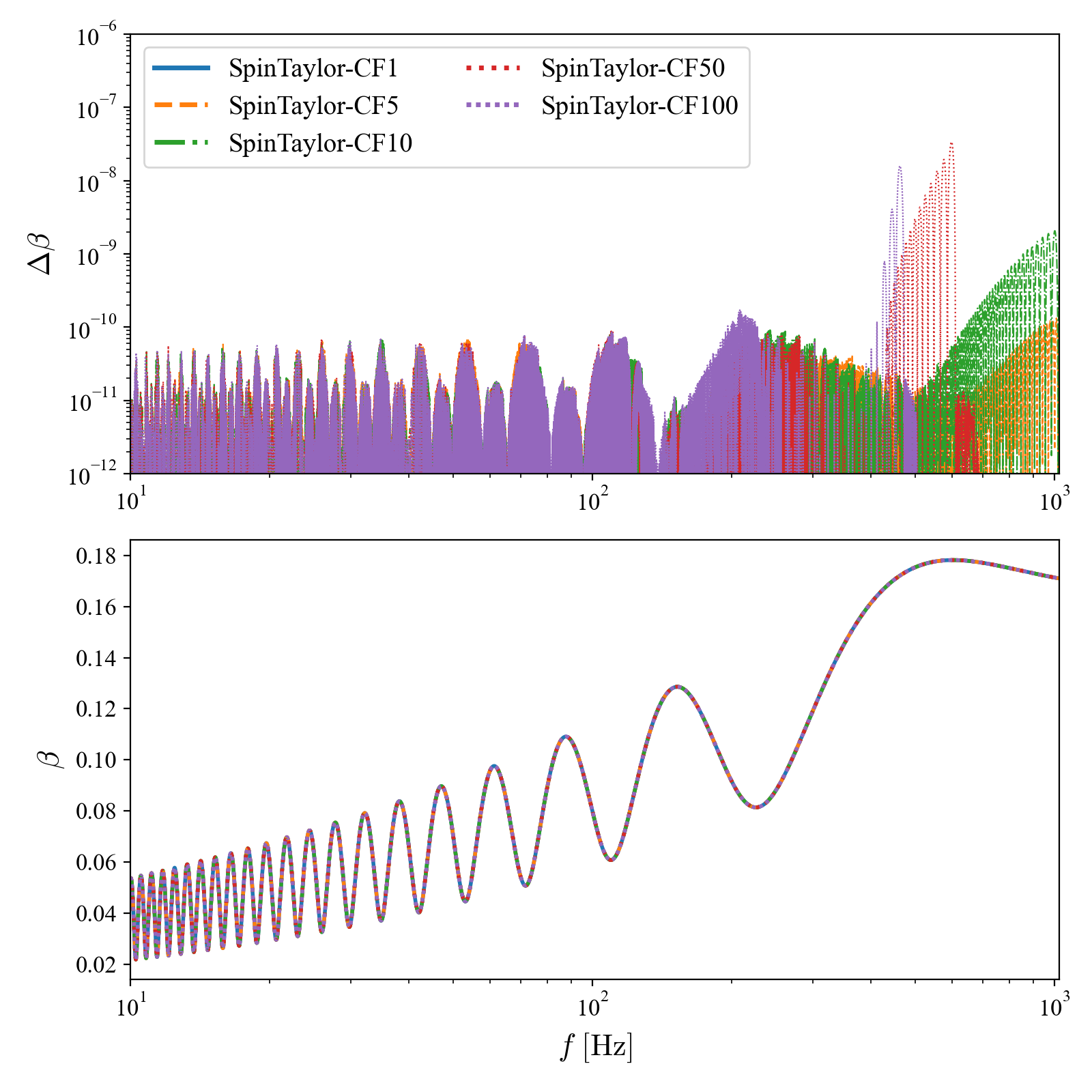}
\caption{The Euler angle $\beta$ returned by the SpinTaylor model for an illustrative BNS system (lower panel; see main text for further details) and the absolute errors with respect to the uniform grid integration, denoted by CF$1$, for a variety of integration steps (upper panel), where CF$n$ corresponds to integration performed with a grid step $n$ times larger in the low frequency range. There is no curve for CF$1$ in the upper panel, since the errors are computed with respect to that case. 
\label{fig:beta_coarse}
}
\end{center}
\end{figure}

In the SpinTaylor version of the model, the integration of the PN spin-precession equations is carried out in the time domain, on two grids: a coarse grid covering the low-frequency part of the inspiral and a fine one covering the high-frequency region. The conversion of the solutions to the frequency domain is performed a posteriori, by mapping the PN velocity parameter $v$ to the gravitational-wave frequency. Hence, the frequency stepping is highly non trivial as, at each time step, one has $f=v^3/(M\pi)$, where $v$ is a time-dependent output of the integration. The transition frequency between the two grids is not fixed, but varies instead with the coarseness of the low-frequency grid: for coarser steps, the fine-grid integration will start earlier, and vice versa, with the exact transition frequency determined by the stopping conditions of the numerical integration and not by a closed form expression. The fine-grid integration is always started a few time steps before the coarse-grid integration stops, to ensure a smooth connection between the two arrays of solution. 
The user can vary the time step used for the coarse-grid integration through the model flag  \textmd{PhenomXPSpinTaylorCoarseFactor}. After the integration, the discrete solutions for the Euler angles $\alpha$ and $\beta$ are expressed as a function of frequency using the stationary phase approximation and subsequently interpolated using cubic splines, whereas $\gamma$ is computed by numerically integrating the minimal rotation condition \cite{Boyle:2011gg}, closely following the implementation in IMRPhenomTPHM~\cite{Estelles:2021gvs}. The lower panel of Fig.~\ref{fig:beta_coarse} shows the $\beta$ angle for an example BNS with mass ratio $1/q=1.5$, total mass $3\,\Msun$ and spins $\chi_1=(0.1,0.2,0.6)$, $\chi_2=(0.4,0.5,0.4)$  at $10$~Hz, with tidal deformabilities following the SLy EOS. 
The upper panel shows instead the absolute differences of $\beta$ for this system evaluated for different choices of the coarse factor, compared to the results for the integration performed using a uniform grid over the full inspiral range, which corresponds to setting the coarse factor to $1$. Differences tend to grow with the integration step size: the growth is more pronounced close to the transition frequency, where the spacing between subsequent frequencies (which is dynamically determined from the evolution of $v$) increases; nonetheless, these differences always remain very small, being $O(10^{-7})$. The absolute errors on $\alpha$ and $\gamma$ stay below $O(10^{-5})$, as is illustrated in Appendix~\ref{app:coarse_angles}.

For this specific binary, the default transition frequency between grids (corresponding to a coarse factor of $10$) is around $1000$~Hz, whereas a coarse factor of $50$ would yield $\sim 600$~Hz (the estimated merger frequency for NRTidalv2 is $\sim 1654$~Hz). Though increasing the coarse factor leads to a lower transition frequency between grids, a coarser low-frequency grid nevertheless yields an overall computational gain (see \ref{subsec:benchmarks}). Together with aggressive multibanding thresholds, the coarse factor option might be useful to reduce the cost of exploratory PE runs \cite{Mateu-Lucena:2021siq}. By default, the code uses a coarse factor of $10$, i.e., the integration at low frequencies is performed with a grid step that is $10$ times larger than the fine grid's step. We will come back to this point in Sec.~\ref{subsec:benchmarks}, where we will show how the waveform evaluation time changes when increasing the coarse grid step.

\section{Model Validation}
\label{sec:tests}

\subsection{Benchmarks}
\label{subsec:benchmarks}

\begin{figure}[h]
\begin{center}
\includegraphics[width=\columnwidth]{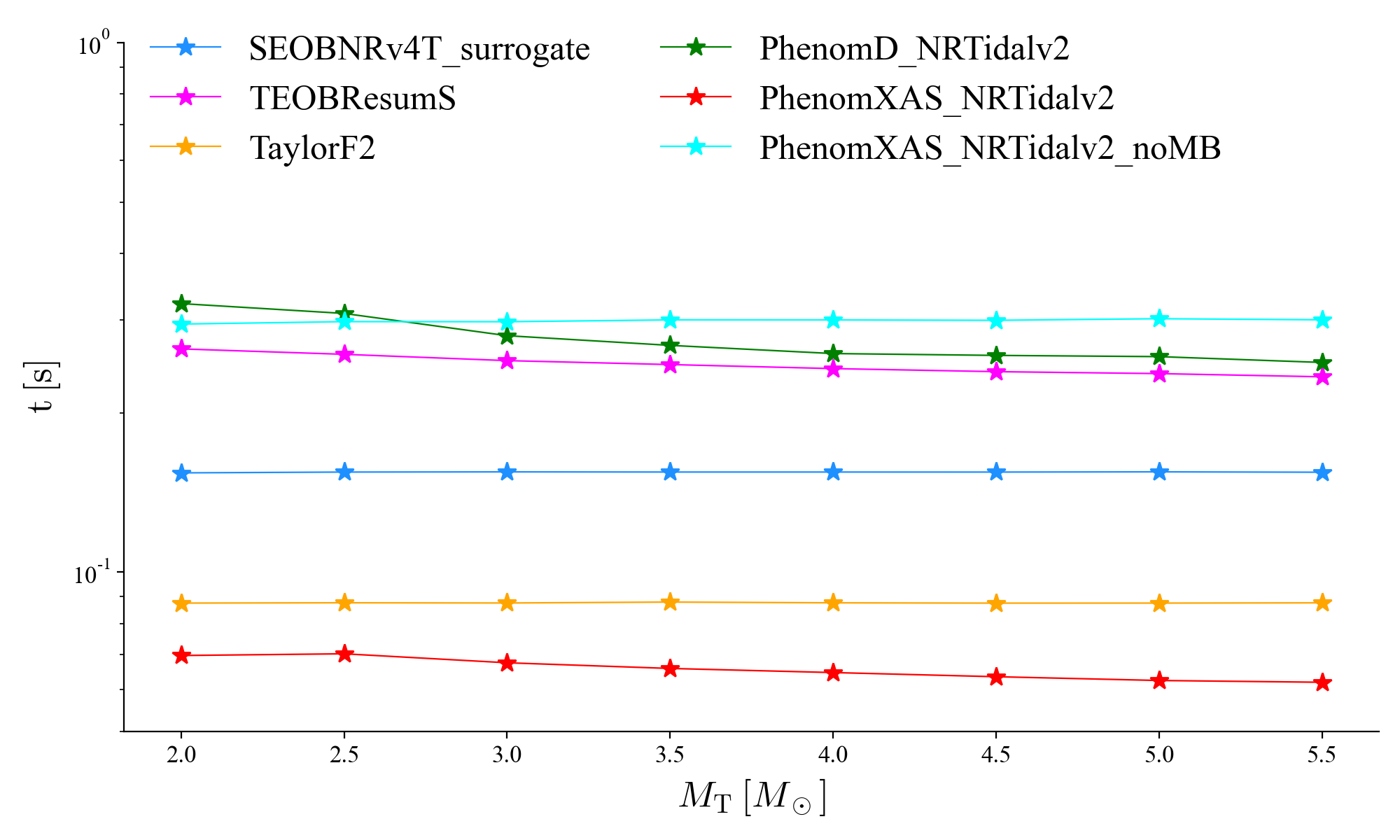}
\caption{A comparison of the evaluation times of several frequency-domain models on a sample of $1000$ random aligned-spin BNS configurations. Each data point indicates the runtimes averaged over a total mass bin of $0.5\,\Msun$.
Timings for \phenxastidal are shown for both the default version and a non-default option (labelled with noMB), where multibanding is deactivated.
\label{fig:timing_bns_aligned}
}
\end{center}
\end{figure}

In this subsection, we compare the evaluation time of \phenxastidal and \phenxptidal to that of other tidal models. Unlike their black-hole counterparts, \phenxas\ and \phenxp, the tidal extensions we describe here come with multibanding activated by default, though the user can switch this off by setting the \textmd{PhenomXHMThresholdMband} and \textmd{PhenomXPHMThresholdMband} flags to zero; just the first for \phenxastidal and both for \phenxptidal. This choice was motivated by the significant improvement in computational efficiency brought by multibanding, which is illustrated in Fig.~\ref{fig:timing_bns_aligned}. In this test, we generate $1000$ random BNS configurations, with component masses between $1 \,\Msun$ and $3\,\Msun$, dimensionless spin magnitudes $\chi_{1,2}\leq 0.5$,\footnote{The upper bounds on $\chi_{1,2}$ was chosen to stay within the validity domain of \textmd{SEOBNRv4T\_surrogate}} and tidal deformabilities uniformly sampled between $0$ and $5000$. We then averaged the evaluation time over bins with a width of $0.5\,\Msun$ in total mass. Each waveform was evaluated with a frequency step of $\Delta f=1/128$~Hz, between $20$~Hz and $2048$~Hz. In the aligned-spin case, we compare \phenxastidal, with and without multibanding, to \textmd{TaylorF2}~\cite{Buonanno:2009zt,Bohe:2013cla,Bohe:2015ana,Yagi:2016bkt,Henry:2020ski}, \phendtidal~\cite{Dietrich:2019kaq}, \textmd{SEOBNRv4T\_surrogate}~\cite{Lackey:2018zvw}, and the frequency-domain version of \teob~v4.1.4 \cite{Damour:2009wj,Bernuzzi:2014owa,Nagar:2018zoe,Gamba:2020ljo}.\footnote{The version of \teob tested here corresponds to the Python package version 0.0.2. For all other waveforms, we use the implementations in {LALSimulation}.} The timing tests have been performed on an Intel Xeon Gold 6148 processor. We can see that the (default) multibanded version of \phenxastidal is the fastest model currently available, being about $4$ times faster than \phendtidal and \teob  (which perform similarly to the non-multibanded version of \phenxastidal), roughly $2$ times faster than \textmd{SEOBNRv4T\_surrogate}, and $\sim 1.3$ times faster than \textmd{TaylorF2}. 

\begin{figure}[h]
\begin{center}
\includegraphics[width=\columnwidth]{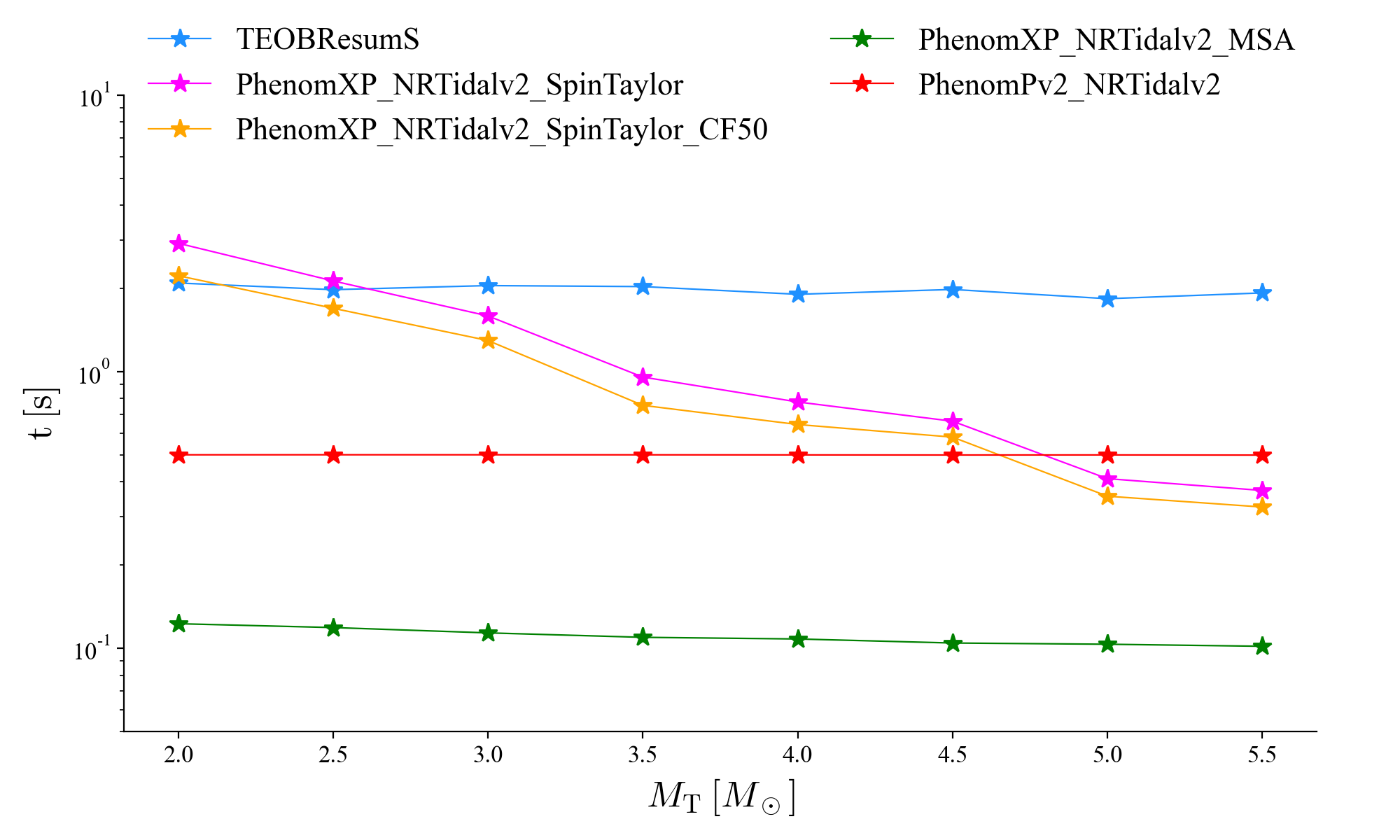}
\caption{A comparison of the evaluation time of several frequency-domain models on a sample of $1000$ random precessing BNS configurations. Each data point indicates the runtimes averaged over a total mass bin of $0.5\,\Msun$. In the case of \phenxptidal, we report the timings for the default version as well as an option with \textmd{PhenomXPSpinTaylorCoarseFactor} set to 50 (labelled as CF50), which is 5 times larger than the default value.
\label{fig:timing_bns_prec}
}
\end{center}
\end{figure}

We repeated a similar test for the precessing approximants, over the same mass range, this time with spins isotropically distributed. For completeness, we present here the timings relative to both the default and the SpinTaylor version of \phenxptidal; for the latter, we consider two values of the coarse factor, the default value ($10$) and $50$. We considered as well \phenptidal and the frequency-domain version of \teob \cite{Gamba:2021ydi,Gamba:2020ljo}. The average runtimes for each model are shown in Fig.~\ref{fig:timing_bns_prec}. Without changing any other aspect of the twisting-up implementation, the numerical evolution of the PN precession dynamics has a large impact on the evaluation time of the model at low masses, due to the increasing computational cost of the integration and of the spline interpolation. However, the evaluation time sharply decreases for heavier binaries, plummeting below that of \phenptidal (\teob) for total masses $M_{\rm T}\gtrsim 5\,\Msun$ ($2.5\,\Msun$). Given the trend of the evaluation time versus total mass,  \phenxptidal-SpinTaylor appears as a viable candidate for future extensions of the model to neutron star--black hole binaries. Increasing the coarse factor by a factor 5 yields an average speedup of $\sim 1.2$ over the mass range considered. \phenxptidal with default MSA angles stands out as the fastest model in the comparison, being $4.5$ times faster than \phenptidal.

\subsection{Comparison to other waveform models}
\label{ssec:model_comparison}

We will quantify the agreement between model (M) and signal (S) by means of the match function:
\begin{equation}
\bar{\mathcal{F}}(h_\mathrm{M},h_\mathrm{S}) = \max_{t_c, \phi_0, \psi_0} \frac{\braket{h_\mathrm{M}, h_\mathrm{S}}}{\sqrt{\braket{h_\mathrm{M}, h_\mathrm{M}}\braket{h_\mathrm{S}, h_\mathrm{S}}}},
\label{eq:mismatch}
\end{equation}
where the overlap is maximised over polarization angle $\psi_0$, coalescence time $t_c$ and reference phase $\phi_0$. In what follows, we will refer to the quantity $1-\bar{\mathcal{F}}$ as the mismatch between two waveforms. For precessing waveform models, we also optimize the match over rigid rotations of the initial in-plane spins, as in~\cite{Pratten:2020ceb}. We use the Advanced LIGO~\cite{TheLIGOScientific:2014jea} design sensitivity Zero-Detuned-High-Power power spectral density \cite{adligopsd} to compute the noise-weighted inner product $\braket{\cdot,\cdot}$ (defined in, e.g.,~\cite{Cutler:1994ys}). Following the match study presented in \cite{Dietrich:2019kaq}, we compute our matches between $f_{\mathrm{min}}=40$~Hz  and $f_{\mathrm{max}}=2048$~Hz.

\subsubsection{Aligned-spin BNS}
\label{subsub:aligned_matches}

\begin{figure}[h]
\begin{center}
\includegraphics[width=\columnwidth]{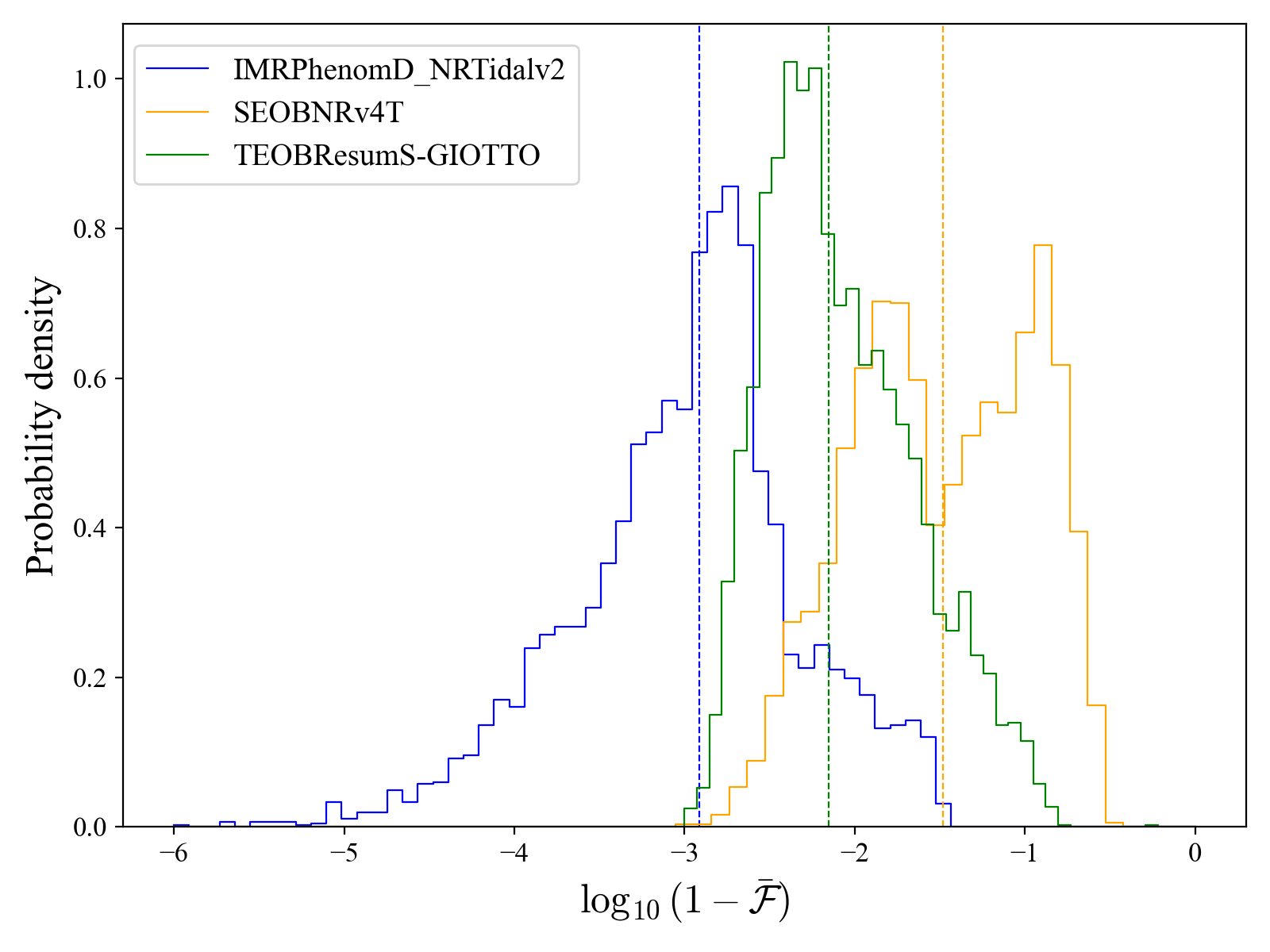}
\caption{A comparison of the distribution of mismatches between \phenxastidal and other aligned-spin models including tidal effects.  \textmd{SEOBNRv4T} is the model showing the largest differences with respect to other models. Vertical dashed lines mark the medians of the three distributions. 
\label{fig:matches_bns_aligned}
}
\end{center}
\end{figure}

\begin{figure}[htbp]
\begin{center}
\includegraphics[width=0.8\columnwidth]{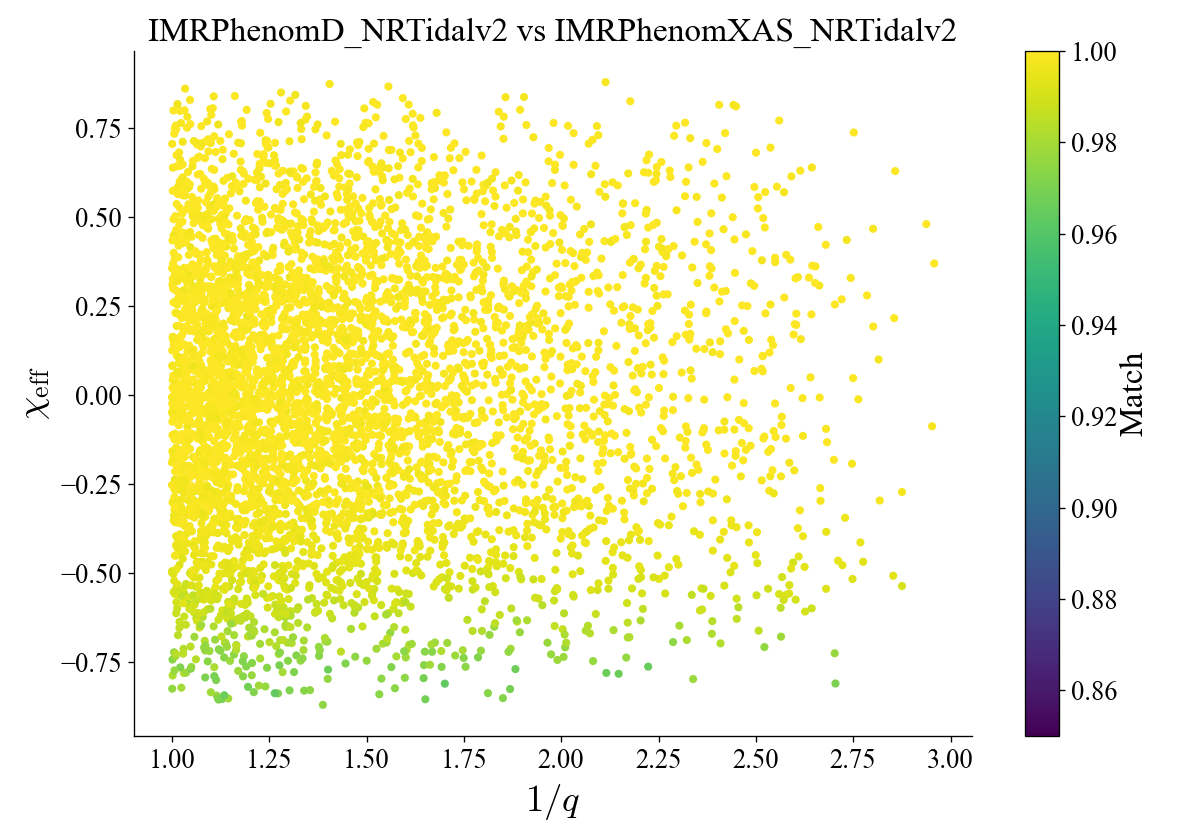}\\
\includegraphics[width=0.8\columnwidth]{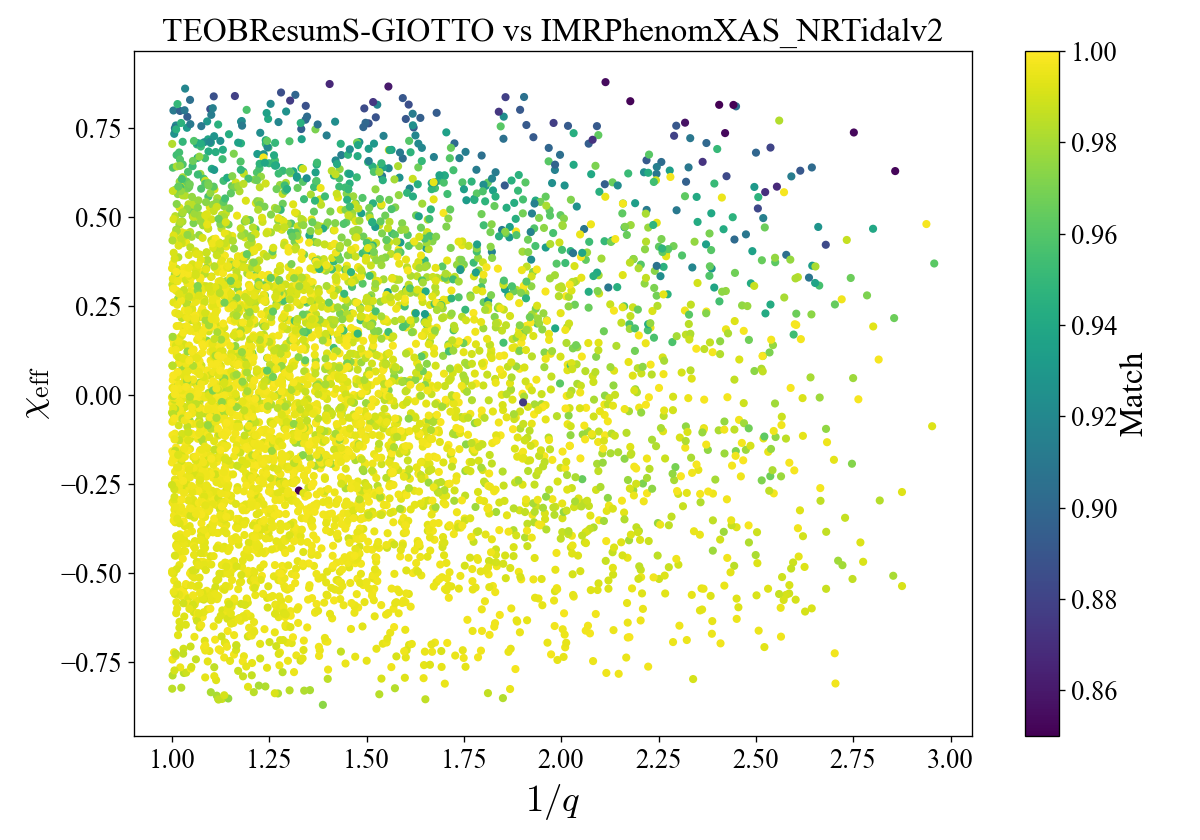}\\
\includegraphics[width=0.8\columnwidth]{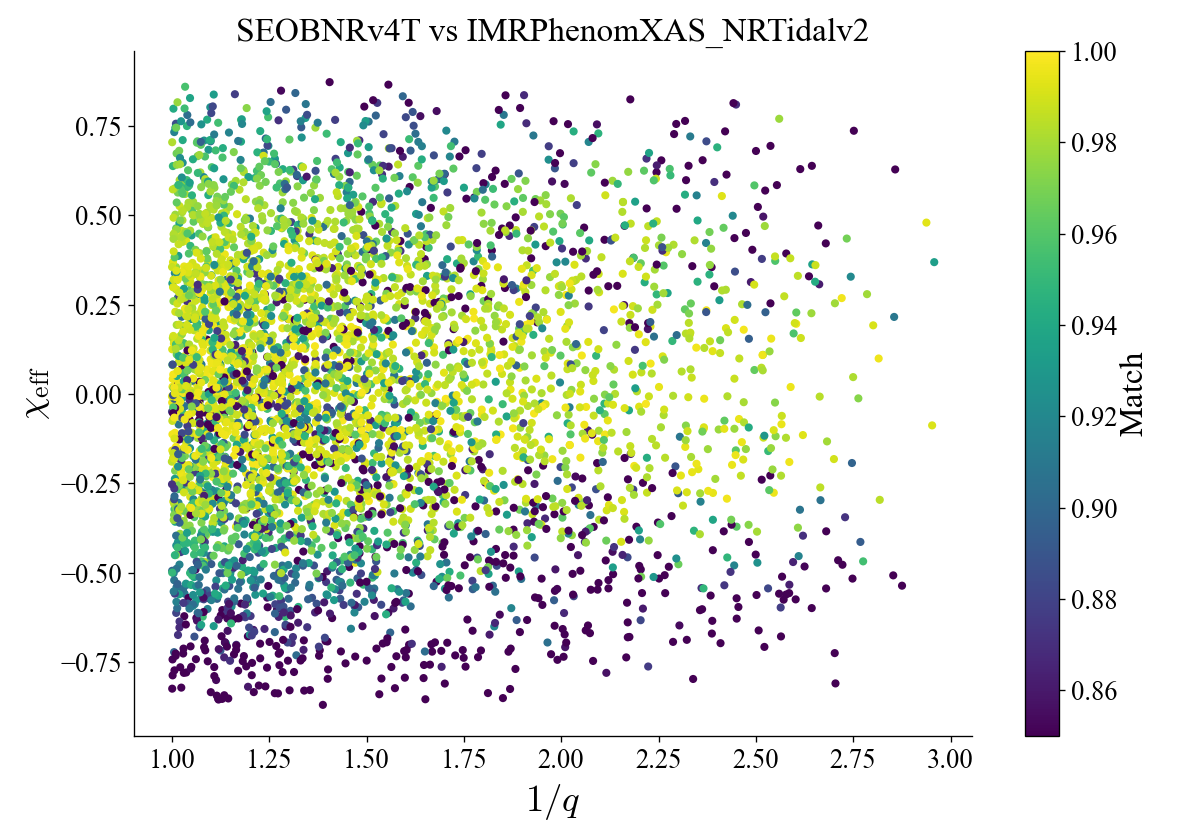}\\
\includegraphics[width=0.8\columnwidth]{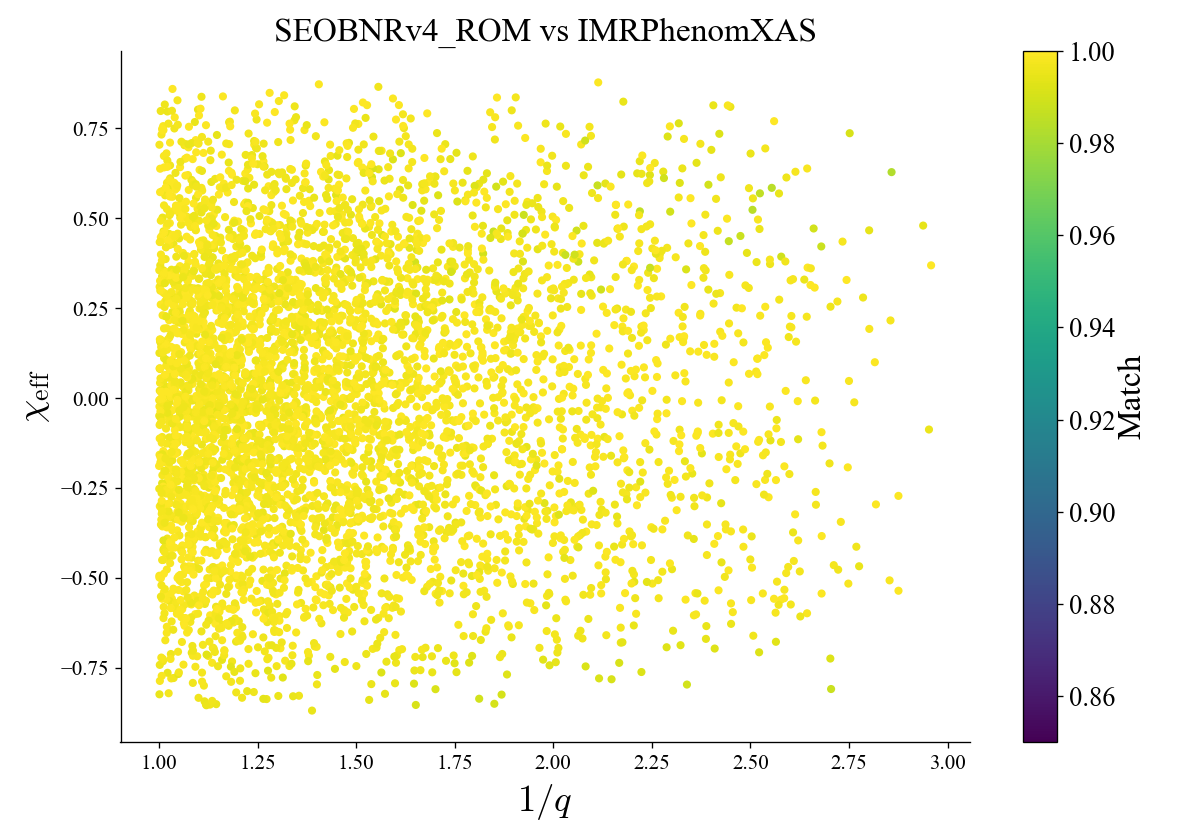}
\caption{2D scatter plots for the match between \phenxastidal and other tidal aligned-spin models, as a function of inverse mass ratio and effective spin. Darker (lighter) points indicate worse (better) agreement between the two models reported in the plot's title. The bottom panel shows a comparison over the same region of parameter space between the BBH models \phenxas and \textmd{SEOBNRv4\_ROM}. This plot suggests that the discrepancies between \textmd{SEOBNRv4T} and \phenxastidal are mostly due to different treatments of tidal effects. \label{fig:2D_aligned_mismatch.png}
}
\end{center}
\end{figure}

Here we compare the aligned-spin model  \phenxastidal  to other models including matter effects, i.e., \phendtidal, the time-domain version of \teob, and \textmd{SEOBNRv4T}~\cite{Hinderer:2016eia,Steinhoff:2016rfi}. We generate $5000$ random aligned-spin binary neutron star configurations, with component masses between $1\,\Msun$ and $3\,\Msun$, dimensionless spin magnitudes for both components uniformly distributed between $0$ and $0.89$, and tidal deformabilities uniformly sampled between $0$ and $5000$. These values were chosen to match the bounds of the ``high-spin'' prior employed in LIGO-Virgo-KAGRA analyses of BNSs (e.g., \cite{LIGOScientific:2017vwq}). As illustrated in Fig.~\ref{fig:matches_bns_aligned}, we observe an excellent agreement between the two \textmd{NRTidalv2} phenomenological models, with a median mismatch of $0.1\%$. There is also a generally good agreement with \teob (with a median mismatch of $0.7\%$), whilst more substantial differences can be seen in the comparison with \textmd{SEOBNRv4T}, for which the median mismatch is significantly higher ($3.3\%$). When restricting to the subsample where $\Lambda_1<\Lambda_2$, as is expected for BNSs (with only possible small deviations due to phase transitions---see, e.g., the discussion in footnote~9 of~\cite{Dietrich:2019kaq}), since we assume $m_1>m_2$, the median mismatches drop by roughly $1\%$. The maximum mismatches against \phendtidal, \textmd{SEOBNRv4T} and \teob are $3.7\%$, $31.6\%$, and $59.3\%$; restricting to $\Lambda_1<\Lambda_2$ does not affect the result relative to \phendtidal, but reduces the maximum mismatches against the other two models to $29.1\%$ for \textmd{SEOBNRv4T} and $14.7\%$ for \teob. The mismatch distribution against \textmd{SEOBNRv4T} is clearly bimodal: the peak corresponding to the worst matches is strongly correlated with antialigned spins, as shown in Fig.~\ref{fig:2D_aligned_mismatch.png}, where we plot the match as a function of mass ratio $q$ and effective spin
\begin{equation}
\chi_{\rm{eff}}=\frac{m_{1}\chi_1^{z}+m_{2}\chi_2^{z}}{m_{1}+m_{2}},
\end{equation}
where $\chi_{1,2}^{z}$ are the dimensionless spin components parallel to the orbital angular momentum.
This might be due to the different treatment of the tidal sector implemented in the \textmd{SEOBNRv4T} model, especially as far as dynamical tidal effects are concerned. We find a similar disagreement between the two models for negative values of $\chi_{\rm{eff}}$ in our injection studies (see also Sec.~\ref{subsec:injections}). To check whether the disagreement might be due to the different binary black hole (BBH) baselines of the two models, we compare SEOBNRv4\_ROM~\cite{Bohe:2016gbl} and IMRPhenomXAS in the lower panel of Fig.~\ref{fig:2D_aligned_mismatch.png}, finding excellent agreement between the two models. 

\subsubsection{Precessing BNS}
\label{subsec:precessing_bns}

\begin{figure}[h]
\begin{center}
\includegraphics[width=\columnwidth]{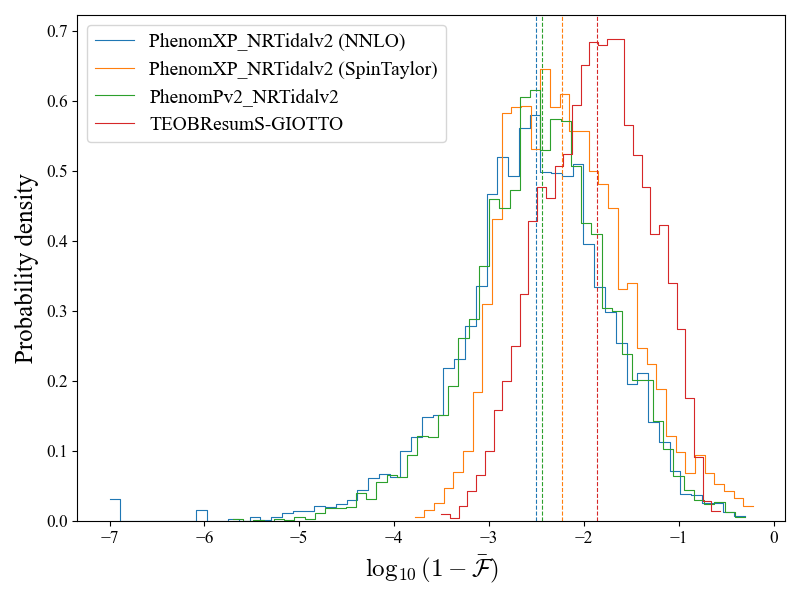}
\caption{The distribution of the mismatch between \phenxptidal, with default MSA angles, and alternative precession prescriptions (blue and orange curves) as well as \phenptidal (green curve). The NNLO version of \phenxptidal implements the same single-spin description employed in \phenptidal. \label{fig:matches_bns_precessing}}
\end{center}
\end{figure}

We now turn to comparing different precessing models. In particular, we are interested in estimating the importance of adding double-spin effects and of going beyond the MSA description. Once again, we generated a sample of $5000$ random precessing BNS configurations, with component masses between $1\,\Msun$ and $3\,\Msun$, spins isotropically distributed and dimensionless spin magnitudes of at most $0.89$. The tidal deformabilities were randomly drawn from a flat prior $\Lambda_{1,2}\in[0,5000]$. 
For each configuration in this sample, we computed the mismatch between \phenxptidal, with default MSA angles, and 1) \phenxptidal with NNLO angles, 2) \phenxptidal with SpinTaylor angles, 3) \phenptidal, 4) the frequency-domain version of \textmd{TEOBResumS-GIOTTO}, which uses a numerical solution of the BBH spin precession equations in the twisting up.\footnote{The version of \textmd{TEOBResumS-GIOTTO} used here (from \url{https://bitbucket.org/eob_ihes/teobresums/pull-requests/8}) includes the $m=0$ modes in the inertial frame polarizations and allows us to only select the $\ell = |m| = 2$ coprecessing frame modes in order to make a direct comparison with \phenxptidal.} The results are shown in Fig.~\ref{fig:matches_bns_precessing}, where dashed vertical lines mark the median of each distribution. The NNLO version of \phenxptidal agrees closely with \phenptidal, which is expected since both models implement the same single-spin approximation.  For the sample considered here, the maximum mismatch between the default version of  \phenxptidal and the SpinTaylor (NNLO) version is $\sim 60\%$ ($50\%$); these numbers decrease to $\sim 54\%$ ($39\%$) when restricting to points with $\Lambda_1<\Lambda_2$. The highest mismatches correspond to edge-on or nearly edge-on configurations, which are particularly sensitive to the precession prescription being used. The maximum mismatch against \teob is instead around $27\%$ on the whole sample, and a few percent lower when imposing $\Lambda_1<\Lambda_2$. When comparing \phenxptidal to single-spin approximants, we find that matches below $95\%$ are around $5\%$ of the total sample and the median mismatch is $\sim 0.3\%$. When comparing the MSA and SpinTaylor versions of \phenxptidal, the percentage of matches below $95\%$ rises to around $9\%$ and the median of the mismatch distribution is slightly shifted towards higher values of mismatch $\sim 0.6\%$. The comparison against \textmd{TEOBResumS-GIOTTO} returns the highest median mismatches. Nevertheless, the median mismatch is still quite small, being only slightly above $1\%$ with roughly $16\%$ of matches below $95\%$.

\begin{figure*}[htbp]
\begin{center}
\includegraphics[width=0.8\columnwidth]{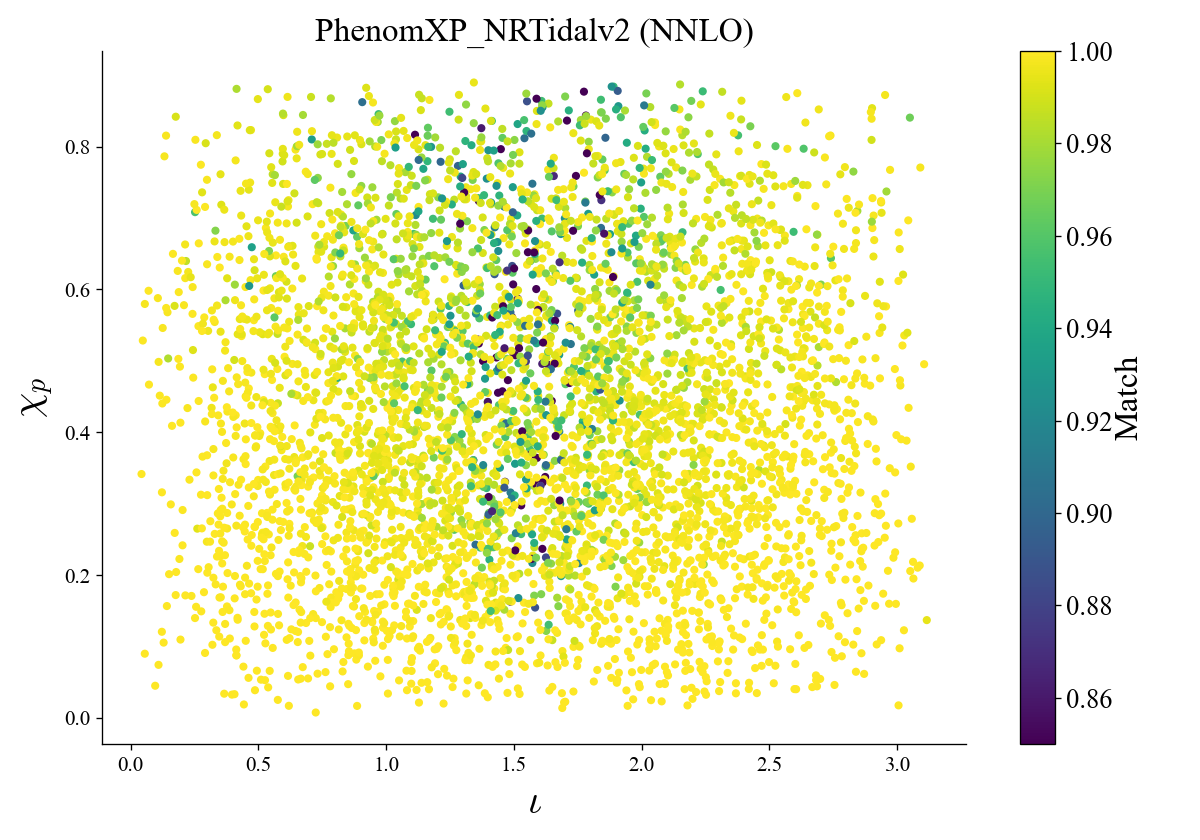}
\includegraphics[width=0.8\columnwidth]{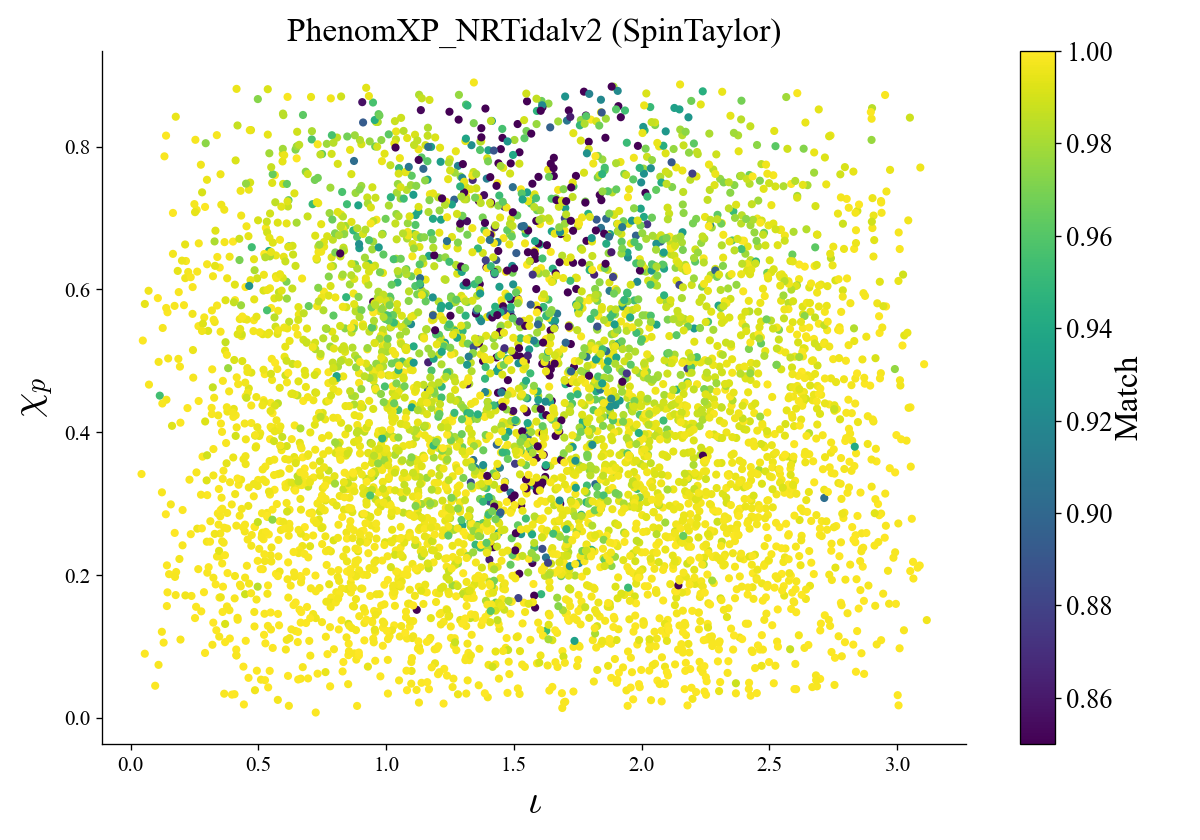}\\
\includegraphics[width=0.8\columnwidth]{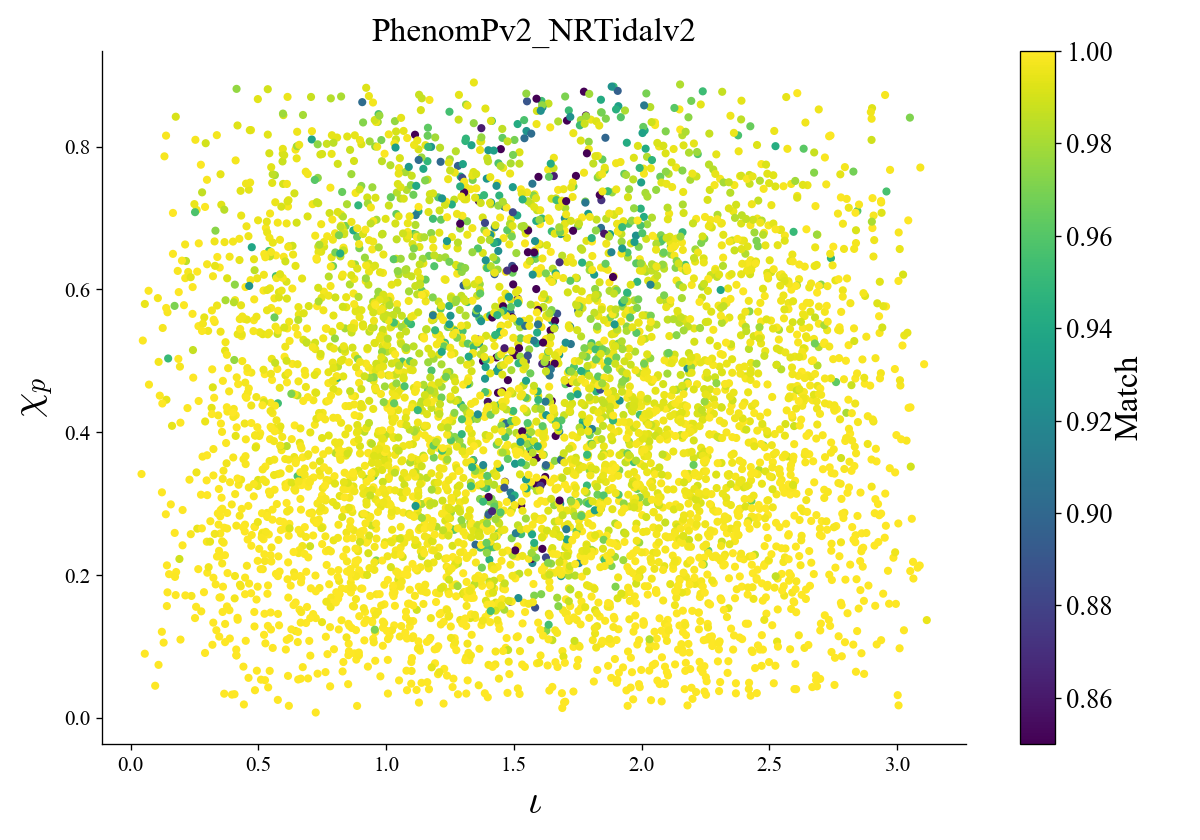}
\includegraphics[width=0.8\columnwidth]{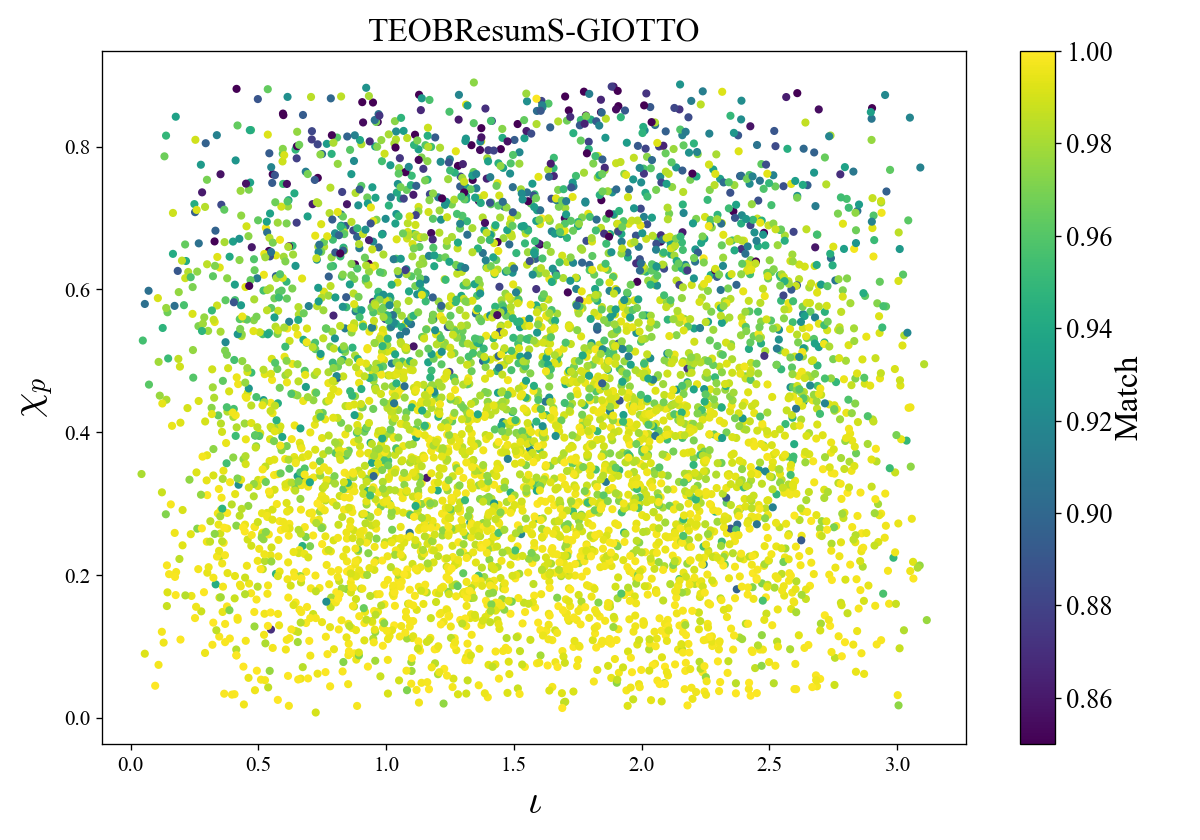}
\caption{The dependence of the matches between different models on the inclination $\iota$ and effective precessing spin parameter $\chi_p$. Top panels: 2D scatter plots for the match between \phenxptidal with MSA angles and the same model with different precession prescriptions, indicated in the titles. Bottom-left(right) panel: 2D scatter plot for the match between \phenxptidal  and \phenptidal (\teob). Alternative descriptions of precession effects become more distinguishable when the source is inclined and spin effects are non-negligible, as can be inferred by the clustering of dark spots in these regions of parameter space. \label{fig:2D_precessing_mismatch.png}
}
\end{center}
\end{figure*}

Fig.~\ref{fig:2D_precessing_mismatch.png} shows the distribution of matches across parameter space. As expected, changes to the precession prescriptions tend to be more relevant when the source has an inclination angle closer to edge-on ($\pi/2$) and a non-negligible effective precessing spin parameter $\chi_p$~\cite{Hannam:2013oca, Schmidt:2014iyl}
\begin{equation}
\chi_p=\frac{1}{A_{1}m_{1}^2}\max{(A_{1}S_{1\perp},A_{2}S_{2\perp})}, 
\end{equation}
with $A_1=2+ 3q/2,A_2=2+ 3/(2q)$ and $S_{1,2\perp}$ denoting the dimensionful spin components perpendicular to the orbital angular momentum. This can be inferred from the clustering of darker spots (denoting poorer matches) around $\iota=\pi/2$ and $\chi_p\gtrsim0.2$. In the comparison against \teob, the mismatch against \phenxptidal grows more uniformly as $\chi_p$ increases, with a weaker dependence on the inclination. This is consistent with our previous comparisons (see Fig.~\ref{fig:2D_aligned_mismatch.png}), where we found that the aligned-spin baselines of the two models already show some non-negligible differences for mild to high values of $\chi_{\rm{eff}}$, which are appreciable even for moderate inclinations in the precessing case. We also observe that all models tend to agree well for low-spin configurations. This will be confirmed by our PE studies (see Sec.~\ref{subsec:gw170817}).

\subsection{Comparison with NR data}
\label{subsec:nr_comparison}

In addition to the comparison with respect to other models, we also want to compare the performance of the waveform model with respect to numerical-relativity waveforms. In this subsection, we separately consider aligned-spin and precessing BNS configurations.

\subsubsection{Aligned-spin binaries} 

\begin{figure}[htbp]
\begin{center}
\includegraphics[width=0.8\columnwidth]{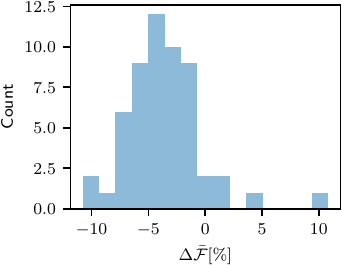}
\caption{Comparison of the relative mismatch between the {\phenxastidal}/{\phendtidal} model and the numerical-relativity simulation data employed in Ref.~\cite{Abac:2023ujg}. One can see a general reduction of the mismatch by (on average) 5\% for the newly implemented \phenxastidal model. We find that \phendtidal performs better only for a very limited number of setups. These setups generally have a small absolute mismatch ($<10^{-3}$). \label{fig:mismatch_hybrids}
}
\end{center}
\end{figure}

We computed the mismatch between \phenxastidal and \phendtidal and the set of hybrids collected in Table~VII of  Ref.~\cite{Abac:2023ujg}, which was used to calibrate the NRTidalv3 model. We use an initial frequency of $350$ Hz to compute the mismatch between the numerical-relativity part of the hybrids and the models, following Eq.~\eqref{eq:mismatch}. The obtained mismatches range between $10^{-3}$ and $10^{-2}$; given that the waveform models considered here only differ by their respective black-hole baseline, and because of the rather moderate mass ratio and spin of the objects, only minor differences are obtained.
These can be better visualised by considering the relative mismatch $\Delta \bar{\mathcal{F}}= 2 (\bar{\mathcal{F}}_{\phenxastidal} - \bar{\mathcal{F}}_{\phendtidal})/
(\bar{\mathcal{F}}_{\phenxastidal} + \bar{\mathcal{F}}_{\phendtidal})$, which is plotted in Fig.~\ref{fig:mismatch_hybrids}. Negative values of $\Delta \bar{\mathcal{F}}$ indicate the superiority of {\phenxastidal}. 
Indeed, for most cases, we find generally a reduction of the mismatch of about 5\%. 

\subsubsection{Precessing binaries} 
\label{ssec:td_comp}

Here we compare \phenptidal and  \phenxptidal to NR simulations of precessing BNS. We will first show some visual time-domain comparisons and will then present a mismatch calculation.
\paragraph{Time-domain comparisons}

We start with some visual comparisons between various phenomenological gravitational waveforms and BAM~\cite{Bruegmann:2006ulg,Thierfelder:2011yi,Dietrich:2015iva,Bernuzzi:2016pie} NR simulations. We pick two illustrative examples from the CoRe database~\cite{Gonzalez:2022mgo}, namely BAM:0142 and BAM:0143, which correspond to the SLy$^{(\swarrow\searrow)}$ and SLy$^{(\searrow\searrow)}$ configurations discussed in Chaurasia~\emph{et al.}~\cite{Chaurasia:2020ntk} (see Table~I therein). Both configurations have equal gravitational masses of $1.35\,\Msun$ and equal dimensionless spin magnitudes of $0.096$, and also both have $\chi_{\rm{eff}}=-0.0676$ and $\chi_p=0.0676$. However, they have a different relative orientation of the in-plane spins (denoted by the direction of the arrows in the configuration names), leading to qualitatively different features in the precessional dynamics, and thus the waveform. In particular, only BAM:0143 exhibits a clear precessional motion of the orbital angular momentum (see Fig.~1 in Chaurasia~\emph{et al.}). We find that while double-spin models like \phenxptidal can adequately capture the differences between the two cases, the single-spin approximation implemented in \phenptidal leads to some inaccuracies that become more evident when precessional effects are enhanced by the inclination of the source. To demonstrate this, we have compared the time-domain polarizations returned by  \phenptidal and \phenxptidal to the NR polarizations  corresponding to BAM:0142  and BAM:0143 (just including the $\ell = 2$ modes), for an inclination angle of $\pi/2$.  To align the NR and model polarizations, we numerically determined the phase and time shift to be applied to the full model strain by minimizing the sum of the squared differences between NR and model polarizations over the retarded time interval $u\in[4,18]$ ms. This is the same interval used to make the comparison with BAM:0143 and another case with \phenptidal given in Fig.~13 of Chaurasia~\emph{et al.}, though they only align the phase. Also as in Chaurasia~\emph{et al.}, we use a reference frequency of $410$~Hz for the spins when generating the model waveforms. In all cases, we plot \phenptidal alongside the two double-spin precession prescription available in \phenxptidal, i.e., the MSA and SpinTaylor options.

\begin{figure*}[htbp]
\begin{center}
\includegraphics[width=2.\columnwidth]{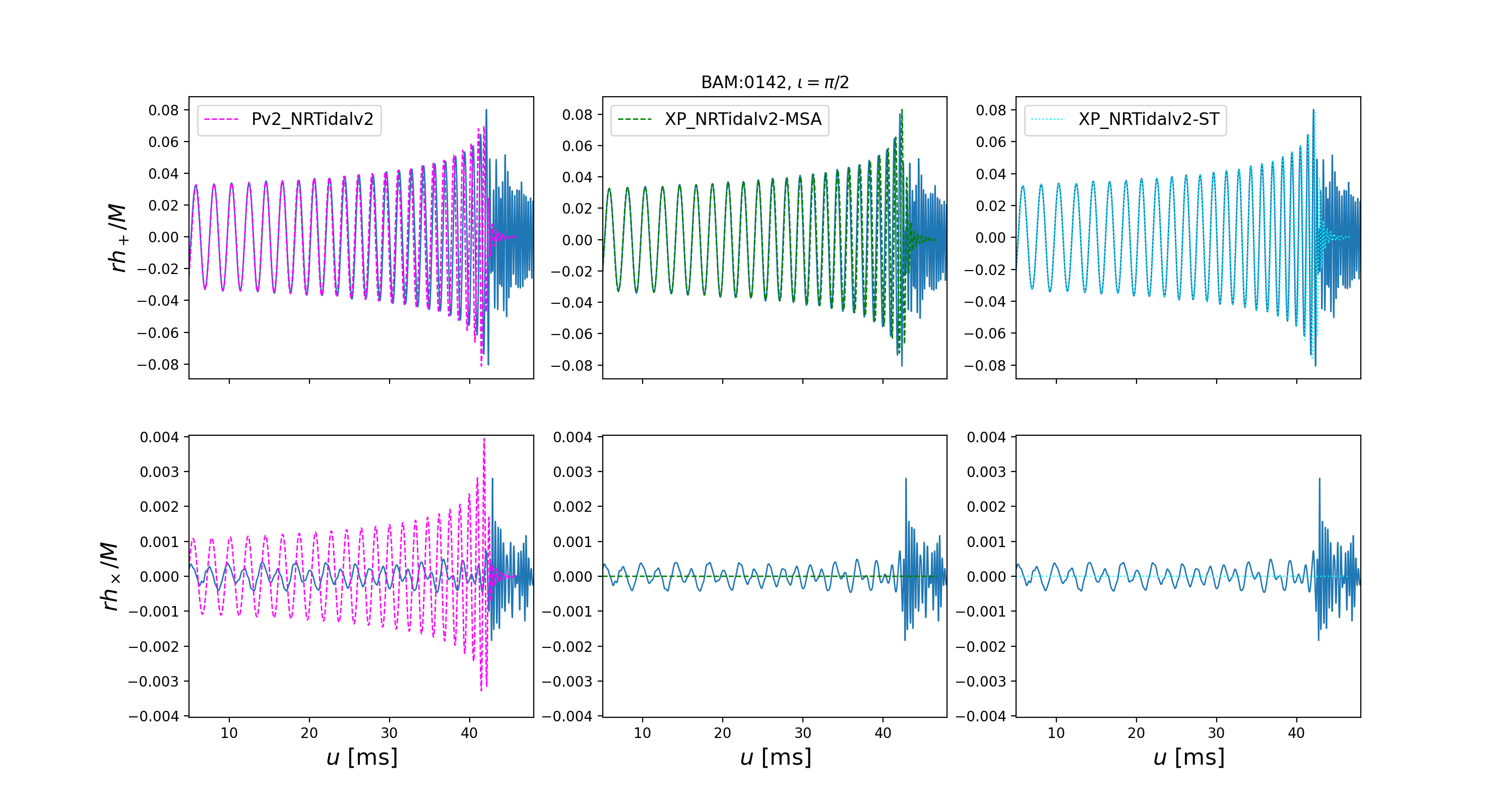}
\caption{Comparison of NR and model predictions for the plus (upper panels) and cross (lower panels) polarizations as a function of retarded time $u$ for a binary with inclination angle $\iota = \pi/2$ and source parameters matching those of the NR simulation BAM:0142. The NR polarizations are shown in blue in all panels, and have the predictions from \phenptidal (magenta, left panels),  \phenxptidal with MSA angles (dark green, middle panels), and  \phenxptidal with SpinTaylor angles (cyan, right panels) superimposed on them. \label{fig:TD_comparison_142_ipi2}
}

\end{center}
\end{figure*}

\begin{figure*}[htbp]
\begin{center}
\includegraphics[width=2.\columnwidth]{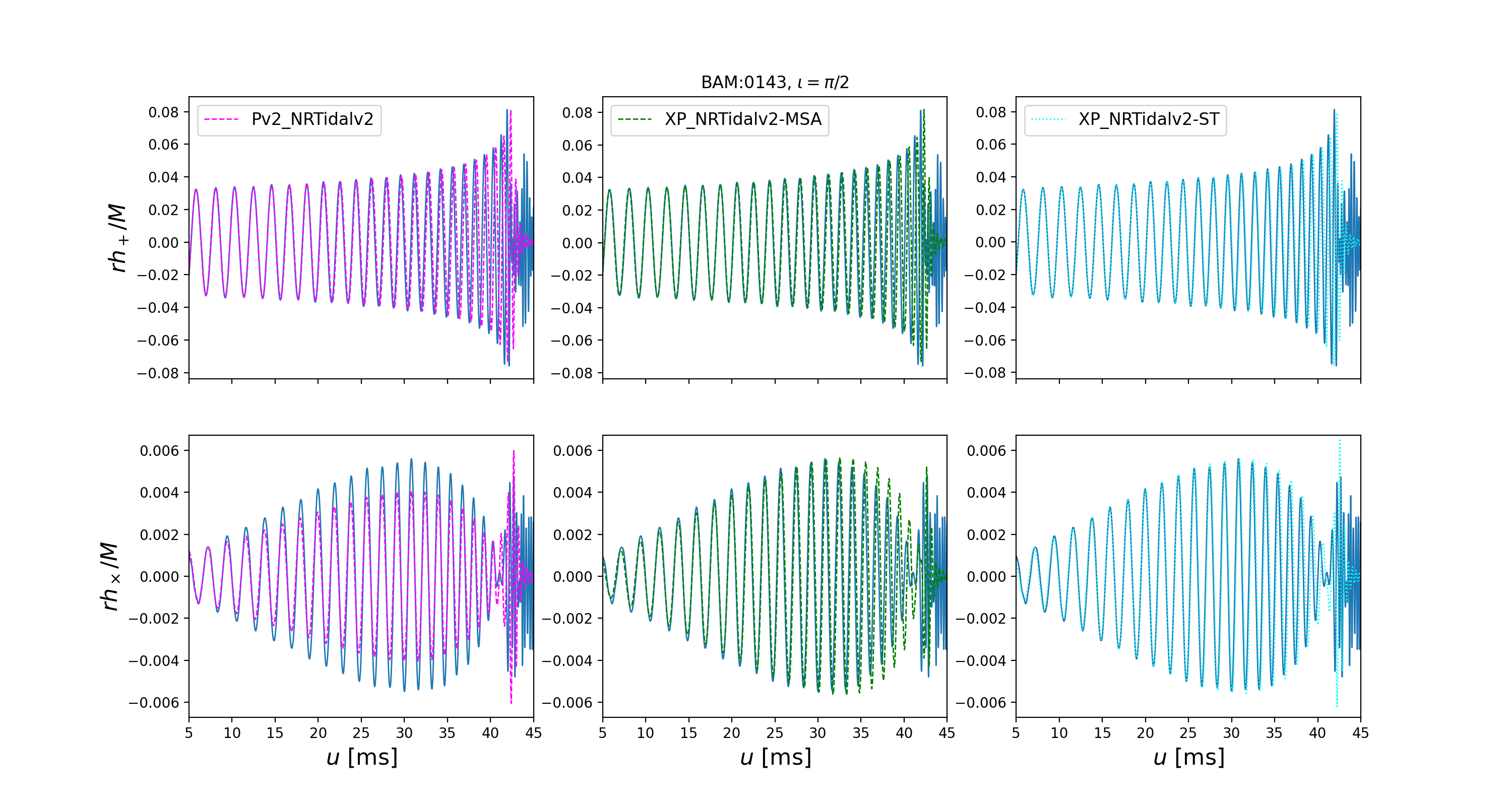}
\caption{Comparison of NR and model predictions for the plus (upper panels) and cross (lower panels) polarizations as a function of retarded time $u$ (given in ms) for a binary with inclination angle $\iota = \pi/2$ and source parameters matching those of the NR simulation BAM:0143. The NR polarizations are shown in blue in all panels, and have the predictions of \phenptidal (magenta, left panels),  \phenxptidal with MSA angles (dark green, middle panels) and  \phenxptidal with SpinTaylor angles (cyan, right panels) superimposed on them.\label{fig:TD_comparison_143_ipi2}
}
\end{center}
\end{figure*}

\begin{figure}[h]
\begin{center}
\includegraphics[width=\columnwidth]{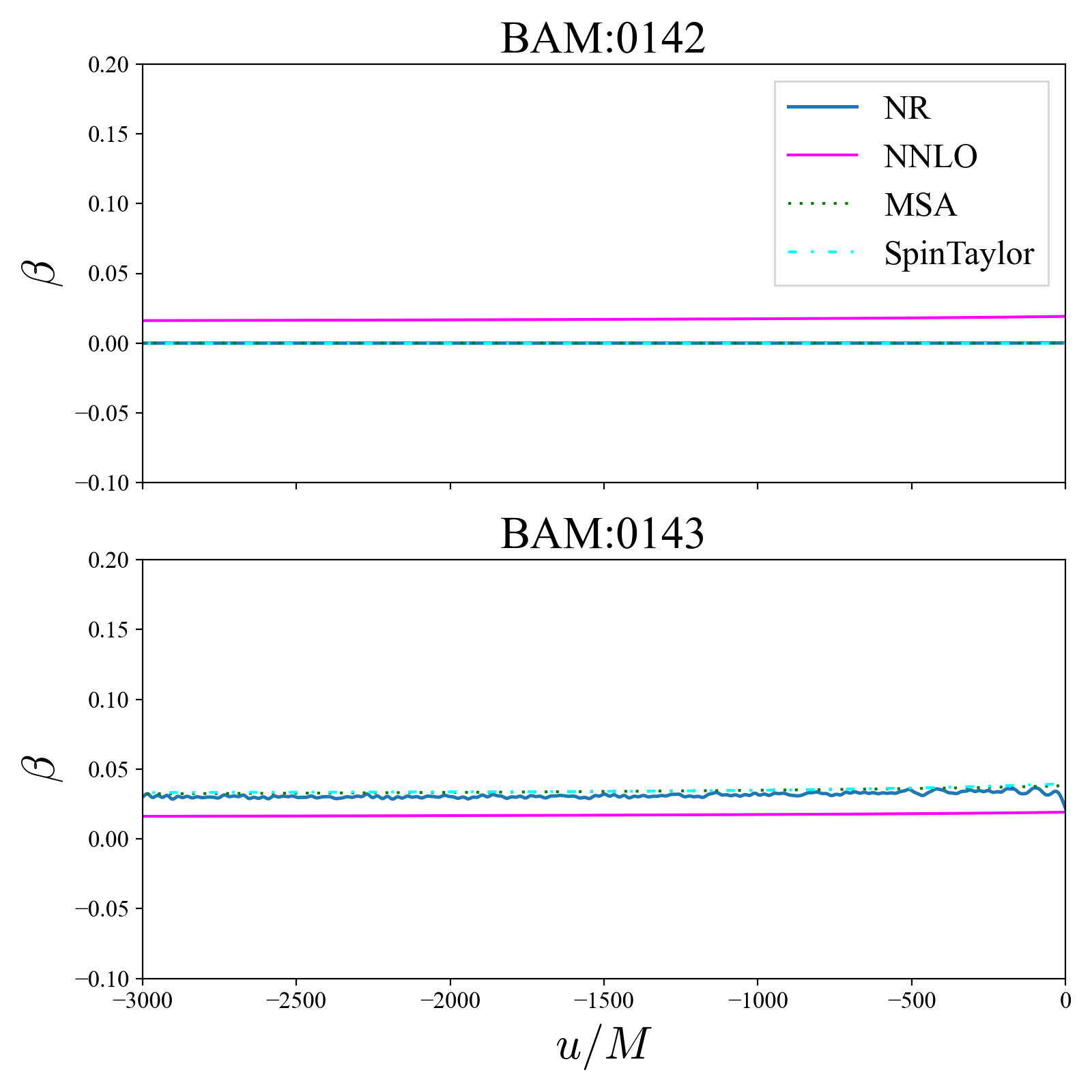}
\caption{Comparison of the evolution of the opening angle of the precession cone as a function of retarded time as computed by \phenptidal (NNLO) and \phenxptidal (MSA and SpinTaylor) for BAM:0142 (upper panel) and BAM:0143 (lower panel). The single-spin approximation employed by \phenptidal does not allow it to distinguish between the two setups, which exhibit qualitatively different dynamics in NR simulations. \label{fig:TD_comparison_beta}
}
\end{center}
\end{figure}

In Fig.~\ref{fig:TD_comparison_142_ipi2}, we show the comparison with BAM:0142. We see that \phenptidal erroneously returns a relatively large signal in the cross polarization, which does not match the NR data as well as \phenxptidal, while the plus polarization is reasonably reproduced by all models.\footnote{We find that the small inspiral signal in the cross polarization from the NR waveforms that is not reproduced by \phenxptidal can be attributed primarily to the asymmetry of $\pm m$ modes due to in-plane spins that is predicted by PN calculations (see, e.g.,~\cite{Boyle:2014ioa}), but is not included in \phenxptidal, since it is not included in the underlying \phenxp model. In particular, we find that if we symmetrize the dominant $h_{2,\pm 2}$ modes by making the replacements $h_{22} \to (h_{22} + h_{2,-2}^*)/2$, $h_{2,-2} \to (h_{22}^* + h_{2,-2})/2$ to enforce the $h_{2,-2} = h_{22}^*$ symmetry present for aligned-spin signals that is used for the coprecessing frame modes of \phenxptidal, the amplitude of the oscillations in the cross polarization inspiral signal is reduced by a factor of $\sim 4$. (Since there is negligible precession of the orbital angular momentum in this case, the inertial frame modes we symmetrize coincide with the coprecessing modes to very good accuracy.)} Fig.~\ref{fig:TD_comparison_143_ipi2} shows a similar comparison, this time for BAM:0143. Here the two \phenxptidal models still better reproduce the amplitude of the signal, especially as far as the cross polarization is concerned; one can also see some differences between the MSA and SpinTaylor prescriptions, with the latter showing a better agreement with NR in the high-frequency part of the inspiral. The differences observed in the GW polarizations can be understood in terms of the evolution of the opening angle of the precession cone, $\beta$, computed by the three models. Fig.~\ref{fig:TD_comparison_beta} shows the evolution of $\beta$ for the two BNS configurations considered here: It is clear that \phenptidal does not distinguish at all between the two cases. The double-spin versions of \phenxptidal, on the other hand, return different results depending on the in-plane spins' configuration and reproduce, in particular, the lack of precessional motion of the orbital angular momentum observed in BAM:0142. We have also checked that switching off matter effects in the twisting-up of the SpinTaylor version leads to a slightly worse agreement with NR data at high frequencies; the version shown here corresponds to the default SpinTaylor version, where such effects are accounted for (see Appendix~\ref{app:settings} for a description of how different settings can be activated). At present, public NR simulations for BNS are relatively short and limited to weakly precessing configurations. Thus, direct comparisons of the Euler angles such as the one presented in Fig.~\ref{fig:TD_comparison_beta} have limited scope: while one can conclude that double-spin waveforms are superior to single-spin ones, it is not possible to clearly show which double-spin approximation is more accurate. This comparison is instead very informative for precessing binary black holes, due to the much wider coverage of parameter space reached by NR simulations: for those systems, the SpinTaylor approximation can be shown to better reproduce the inspiral Euler angles extracted from the simulations~\cite{Colleoni:2024knd}. It will be interesting to perform further comparisons once more simulations of precessing BNS become available.

\paragraph{Mismatches}

\begin{figure}[htbp]
\begin{center}
\includegraphics[width=\columnwidth]{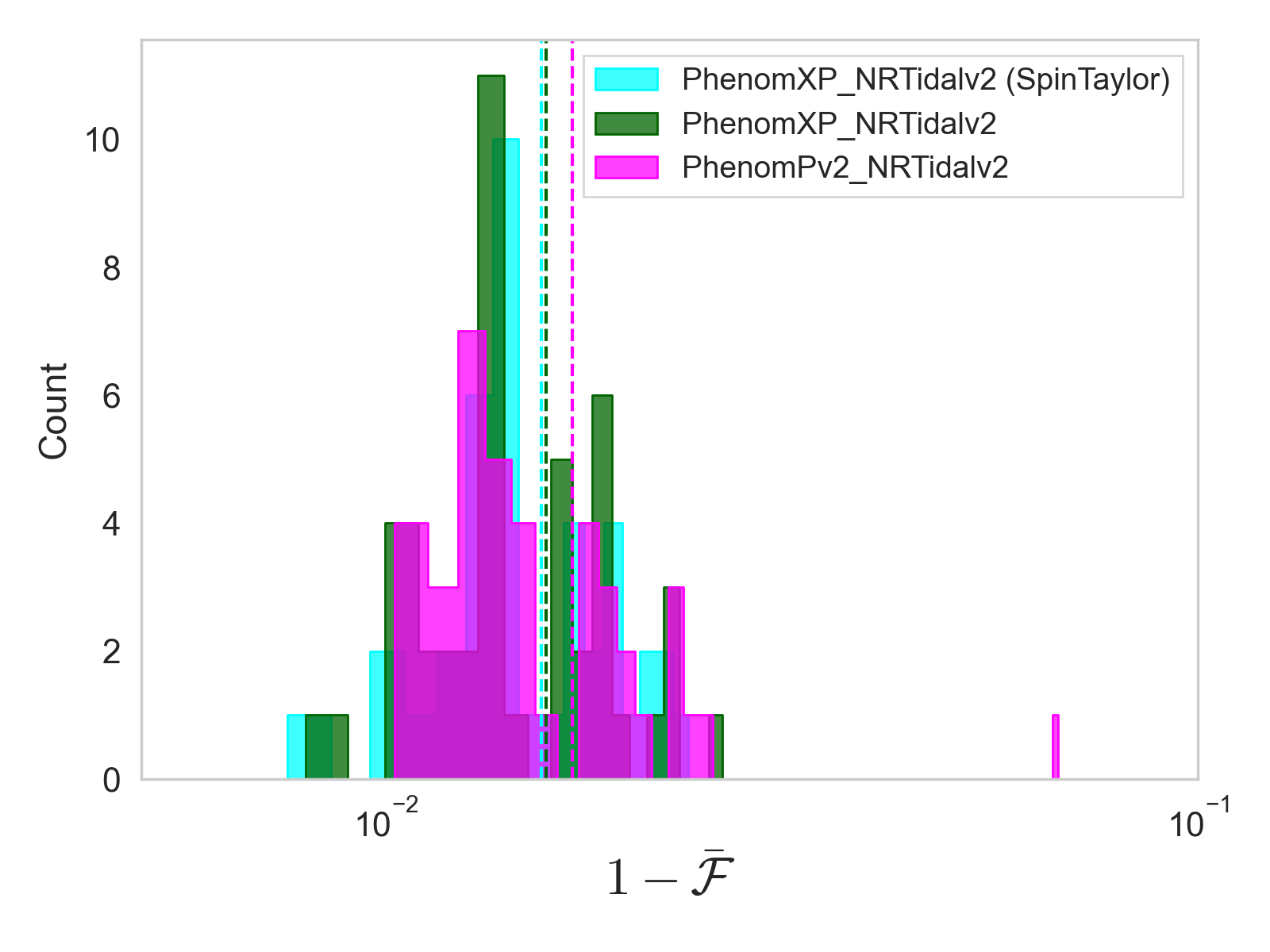}
\caption{Mismatch distributions with respect to the precessing NR simulations of Chaurasia~\emph{et al.}~\cite{Chaurasia:2020ntk}, for a variety of frequency-domain waveform models (magenta: \phenptidal, green: \phenxptidal with MSA angles, cyan: \phenxptidal with SpinTaylor angles). Dashed vertical lines indicate the mean of the distributions. \label{fig:precessing_match_NR}
}
\end{center}
\end{figure}

We further quantify the agreement with the NR simulations of Chaurasia~\emph{et al.}~\cite{Chaurasia:2020ntk} through a computation of the match $\bar{\mathcal{F}}$ of Eq.~\eqref{eq:mismatch}, for a range of values of the source polarization angle $\psi$ and inclination $\iota$. Given that the starting frequency of these simulations is $\sim$407 Hz, we compute the match between 430 and 2048 Hz, maximising over rigid rotations of the in-plane spins and using the Advanced LIGO design sensitivity noise curve. The results are shown in Fig.~\ref{fig:precessing_match_NR}. Once again, we see that differences between models are nearly negligible, though double-spin models offer a modest improvement. This is in line with the mismatch studies discussed in the previous subsections. The simulations considered here have very small effective precessing and aligned spins, as well as equal masses: we do not expect to see significant differences in this region of parameter space, and thus the comparison is not particularly compelling.

\begin{figure*}[htbp]
\begin{center}
\includegraphics[width=2.\columnwidth]{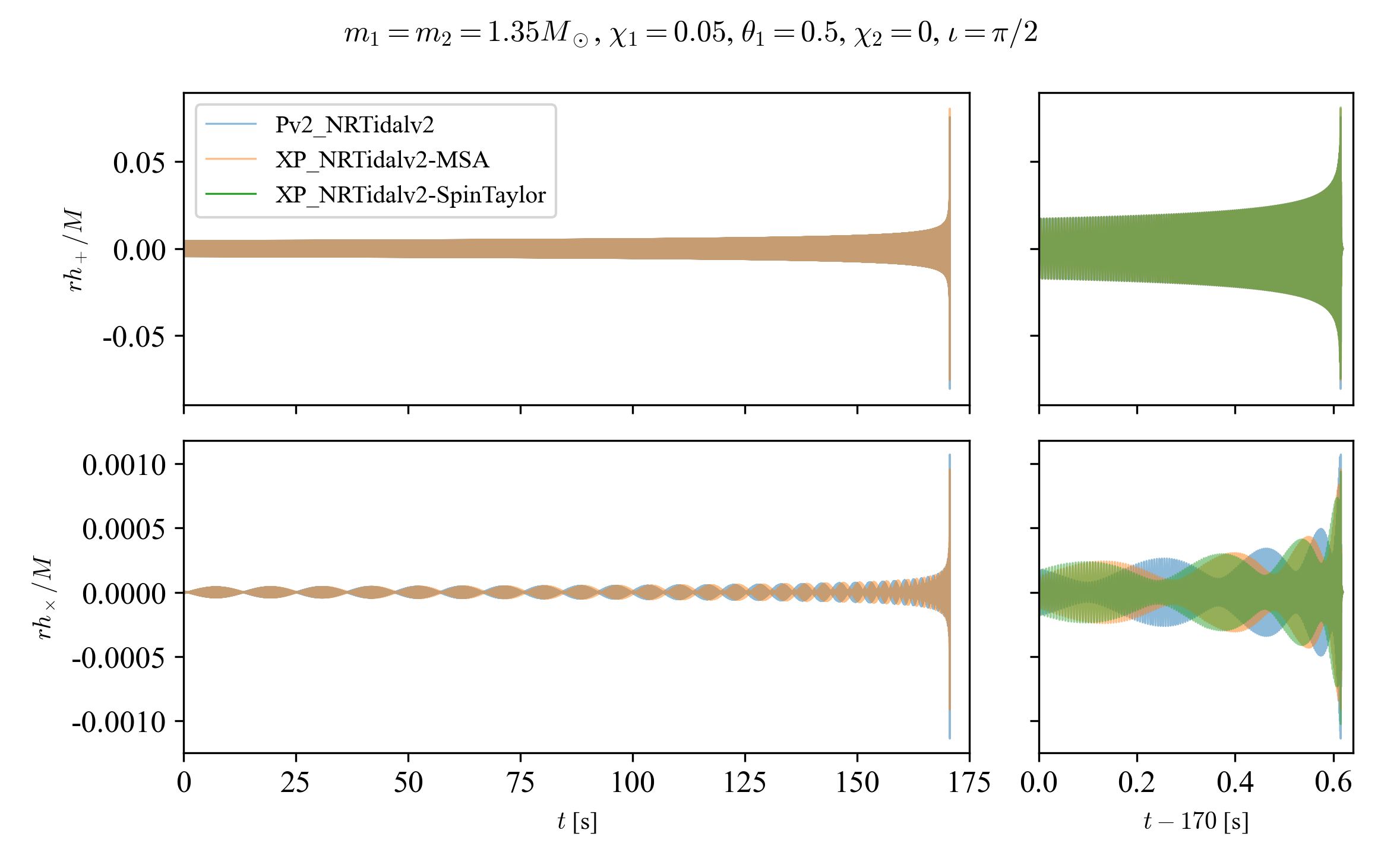}
\caption{Comparison of the plus (upper panels) and cross (lower panels) polarizations for the \phenptidal and \phenxptidal (MSA and SpinTaylor) models using the single-spin configuration described in the text. We plot all three waveforms in the zoomed-in panels on the right, but plot just the \phenptidal and \phenxptidal MSA results in the zoomed-out panels on the left, for clarity. \label{fig:TD_comparison_single_spin}
}
\end{center}
\end{figure*}

While these spin configurations may be astrophysically unlikely for a BNS, we still find that differences in the precession prescription in \phenxptidal lead to changes in the waveform for even single-spin configurations, which Zhu and Ashton~\cite{Zhu:2020zij} argue are astrophysically likely. We illustrate the difference in the time domain in Fig.~\ref{fig:TD_comparison_single_spin} for a case with the same masses and tidal deformabilities as the previous comparisons, but with just one star spinning, with a magnitude of $0.05$ (the upper bound of the low-spin prior for neutron stars used in LVK analyses, e.g.,~\cite{LIGOScientific:2017vwq}), and a relatively moderate misalignment with the orbital angular momentum, with an angle of $0.5$~rad at $20$~Hz. We illustrate the difference in precession prescriptions on the waveform in the time domain in an edge-on case ($\iota = \pi/2$) in Fig.~\ref{fig:TD_comparison_single_spin}. We find that there are noticeable differences between \phenptidal and \phenxptidal in the cross polarization in this case, and even some differences between the MSA and SpinTaylor versions of \phenxptidal. We also find notable differences between the SpinTaylor version with just BBH spin precession and the default one including matter effects, though we do not plot the former, to avoid crowding.

To quantify all these differences, we compute the matches in the cross polarization using the same noise curve and maximizations as in Sec.~\ref{ssec:model_comparison}, though here we compute the matches from $20$~Hz (the high-frequency cutoff is still $2048$~Hz). We find that there is a match of only $49\%$ between \phenptidal and \phenxptidal, $97\%$ between the MSA and SpinTaylor versions of  \phenxptidal, and $79\%$ between the BBH and BNS SpinTaylor precession prescriptions. However, these matches increase considerably to $99.4\%$, $99.95\%$, and $99.8\%$ for $\iota = 0.49\pi$.\footnote{It is surprising that the matches between the two SpinTaylor versions are smaller than those between MSA and (BNS) SpinTaylor. This occurs because the non-BBH contributions make the SpinTaylor result closer to MSA: If one computes the matches between MSA and SpinTaylor for zero tidal deformabilities, then one obtains matches of $58\%$ and $99.6\%$ for $\iota = 0.5\pi$ and $0.49\pi$, respectively.} Thus, these results are in agreement with the results in, e.g., Fig.~\ref{fig:2D_precessing_mismatch.png}, which shows that even though the matches are smaller for close-to-edge-on systems, they are still large for small spins, as we are considering here (since the random sampling is unlikely to have given inclination angles too close to $\pi/2$).

It should be also stressed that the amplitude of the cross polarization is significantly suppressed as the binary approaches an edge-on orientation (becoming roughly two orders of magnitude smaller than that of the plus polarization when $\iota=\pi/2$), and we thus do not expect these differences will have an appreciable impact on the analysis of gravitational-wave events with these parameters at current or near-future detector sensitivities. For instance, if one rotates the polarization angle by $0.1$ away from the value that just gives the pure cross polarization, so that there is a $\sim 10\%$ contribution from the plus polarization, then one obtains a match of $99.93\%$ between \phenptidal and \phenxptidal even for the exactly edge-on case.

\section{Parameter estimation studies}
\label{sec:pe}

\subsection{Real events}

\subsubsection{GW170817}
\label{subsec:gw170817}

GW170817~\cite{LIGOScientific:2017vwq} was the first gravitational-wave signal detected
that likely arose from the coalescence of two neutron stars.
We reanalyse this event with \textmd{IMRPhenomXP\_NRTidalv2} using serial \textmd{bilby} \cite{Ashton:2018jfp} with the \textmd{dynesty} \cite{Speagle_2020} nested sampler and its default settings, analyzing $128$~s of data from LIGO Open Science Center~\cite{LIGOScientific:2019lzm} frames between $23$ and $2048$~Hz. We use \textmd{bilby}'s \textmd{UniformInComponentsChirpMass} and \textmd{UniformInComponentsMassRatio} priors,  with prior bounds $\mathcal{M}\in[1.18,1.22]$, $q\in[0.125,1]$ (with an additional constraint on individual masses $0.5\,\Msun\leq m_{1,2}\leq 4\,\Msun$), and a luminosity distance prior $\propto d_{\mathrm{L}}^2$ with $d_{\mathrm{L}}\in[1,75]$~Mpc, matching~\cite{LIGOScientific:2018hze}.  
We take the components' spins to be isotropically distributed, with their magnitudes restricted to be $\chi_{1,2}\leq0.05$, i.e., the low-spin prior used in LIGO-Virgo analyses of GW170817 (e.g.,~\cite{LIGOScientific:2017vwq,LIGOScientific:2018hze}). We do not fix the right ascension and declination of the source. For the tidal sector, we impose uniform priors on the tidal deformabilities of both components, with $\Lambda_{1,2}\in[0,5000]$. We marginalize over detector calibration uncertainties~\cite{SplineCalMarg-T1400682} employing the uncertainty envelopes from the GWTC-1~\cite{LIGOScientific:2018mvr} public release~\cite{GWTC-1-release}, which also gives the power spectral densities we use. We also compare our results to the posterior samples in this data release. Each of the two seeds had a computational cost of $\sim 9700$~CPU hours. The runs were performed on a single $48$-core node ($2.20$~GHz A64FX CPUs) of the cluster MareNostrum~\cite{MN4} exploiting \textmd{bilby}'s local parallelization functionalities.

\begin{figure*}[h]
\includegraphics[width=0.65\columnwidth]{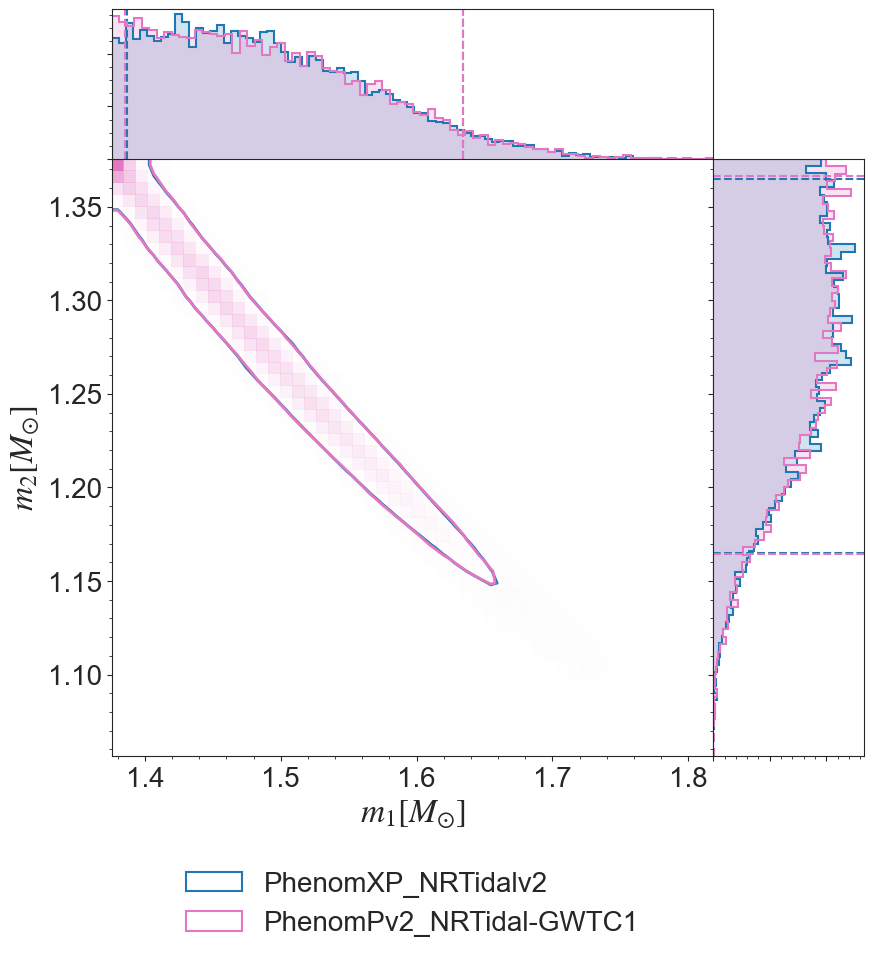}
\includegraphics[width=0.65\columnwidth]{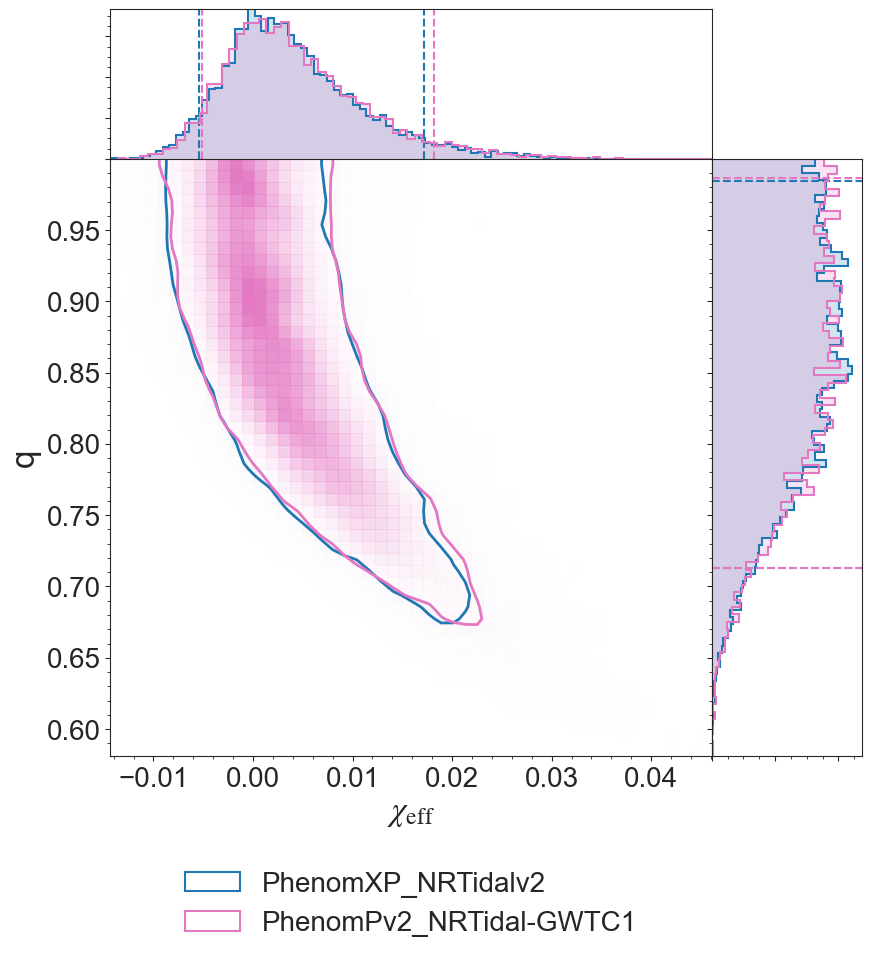}\\
\includegraphics[width=0.65\columnwidth]{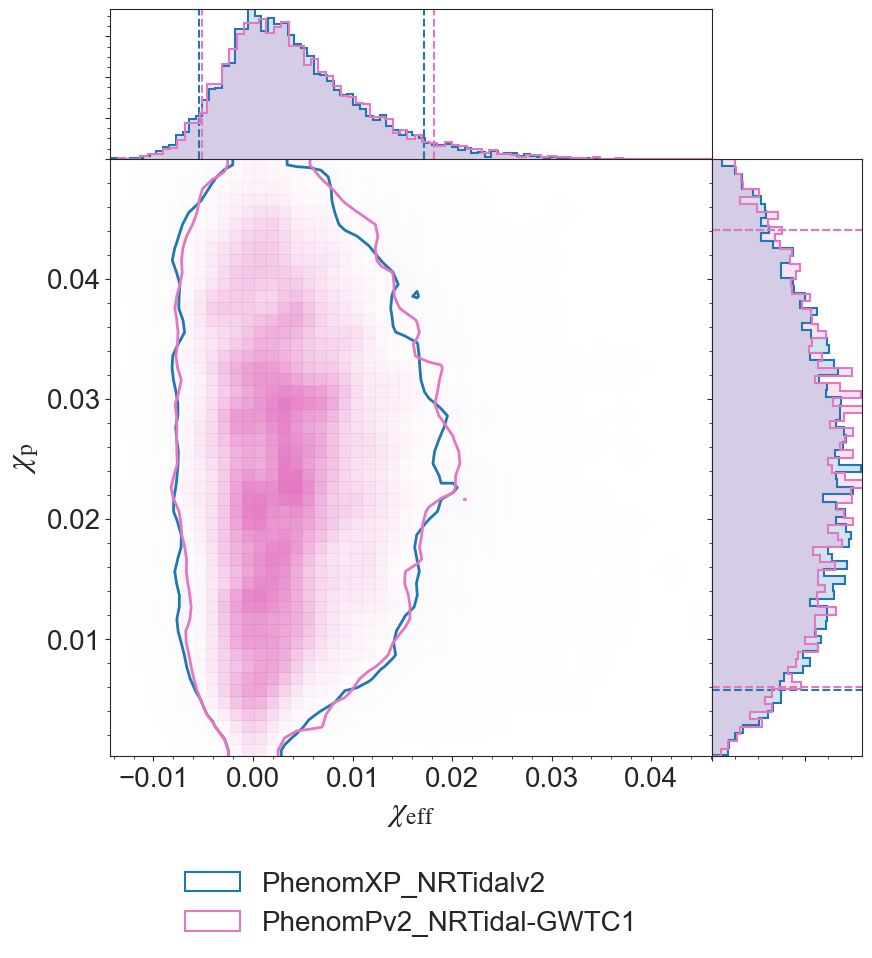}
\includegraphics[width=0.65\columnwidth]{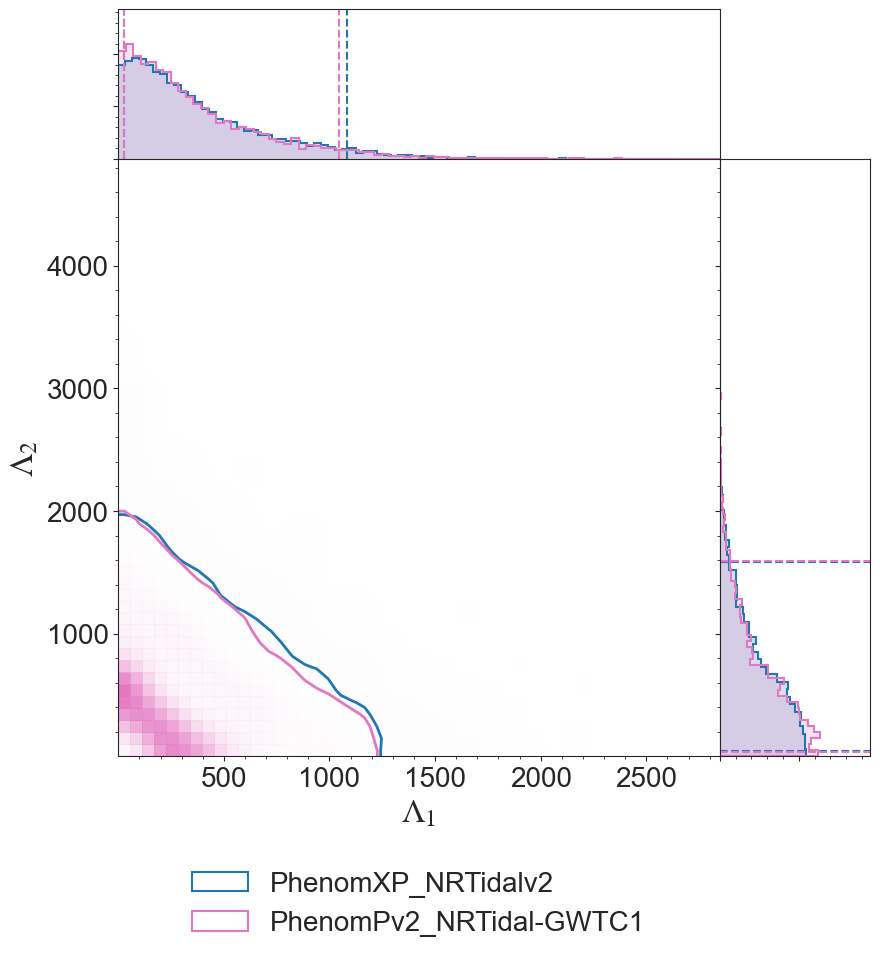}\\
\caption{Posterior distributions obtained from the analysis of GW170817, as estimated using \textmd{IMRPhenomXP\_NRTidalv2} (blue) and \textmd{IMRPhenomPv2\_NRTidal} (magenta, results from GWTC-1~\cite{LIGOScientific:2018mvr}). Dashed (solid) lines mark the $90\%$ credible intervals of 1D (2D) posteriors. The upper panels show detector-frame masses (left), and effective spin and mass ratio (right); the lower panels show the effective and precession spin (left) and the tidal deformabilities of the components (right)\label{fig:pe_gw170817}}
\end{figure*}

The GWTC-1 run analysis employed a different waveform model, \textmd{IMRPhenomPv2\_NRTidal}, and a different sampling method, Markov-chain Monte Carlo sampling, as implemented in LALInference~\cite{Veitch:2014wba}. Despite the different setups of the two analyses, we find excellent agreement between the two sets of results, as shown in Fig.~\ref{fig:pe_gw170817}, in line with the match study of Sec.~\ref{subsub:aligned_matches}. \phenxptidal returns a comparable constraint on the effective spin,
 while $\chi_p$ remains essentially unconstrained, as can be seen in the bottom left panel of Fig.~\ref{fig:pe_gw170817}.

\subsection{Studies of simulated signals}
\label{subsec:injections}

While observations of Galactic binaries indicate that NS in merging binaries should be slowly spinning \cite{Zhu:2017znf,LIGOScientific:2017vwq}, dynamical formation might lead to systems where one or both components have a more significant spin \cite{Zhu:2020zij}, though current calculations~\cite{Ye:2019xvf} suggest that the dynamical channel will not contribute significantly to the rate of BNS mergers. However, it is also possible that the BNS systems observed through gravitational waves are not generated through the same formation channels as observed Galactic BNSs containing pulsars. In particular, the likely BNS GW190425~\cite{LIGOScientific:2020aai} was unexpectedly massive compared to the observed Galactic BNS distribution. Accurate inference of the spin parameters of BNSs is therefore important to be able to identify systems with unexpected properties.

\begin{figure*}[htbp]
\begin{center}
\includegraphics[width=0.65\columnwidth]{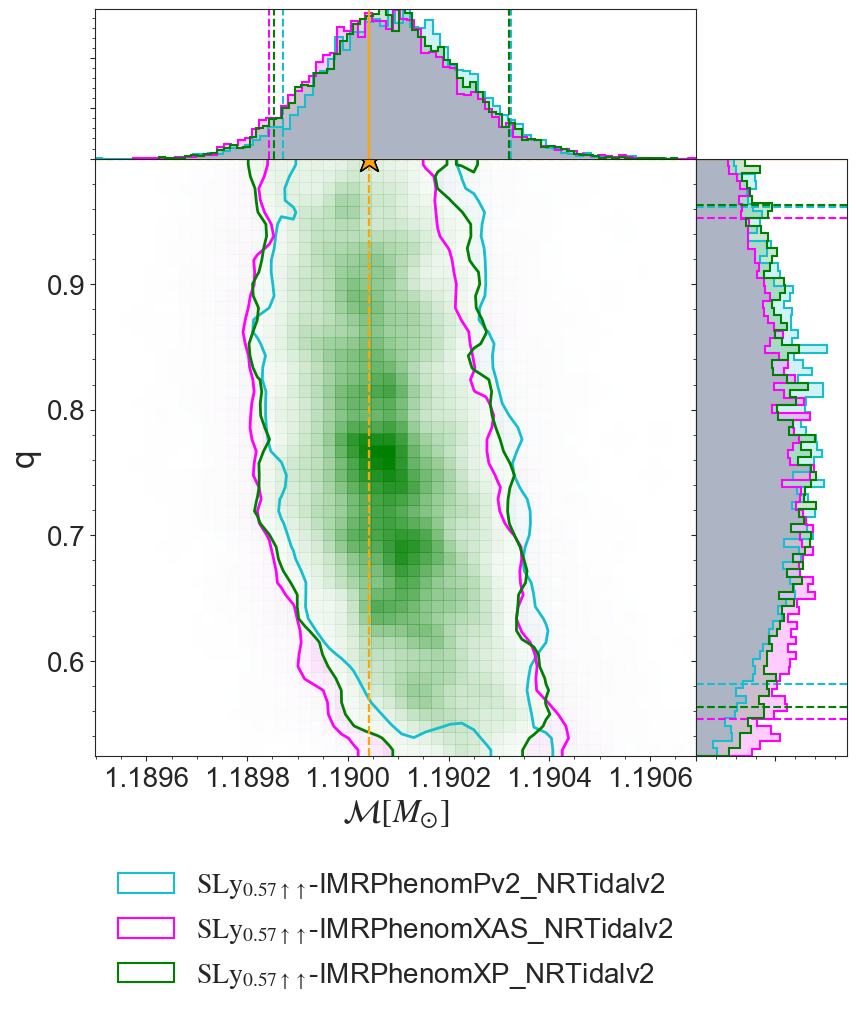}
\includegraphics[width=0.67\columnwidth]{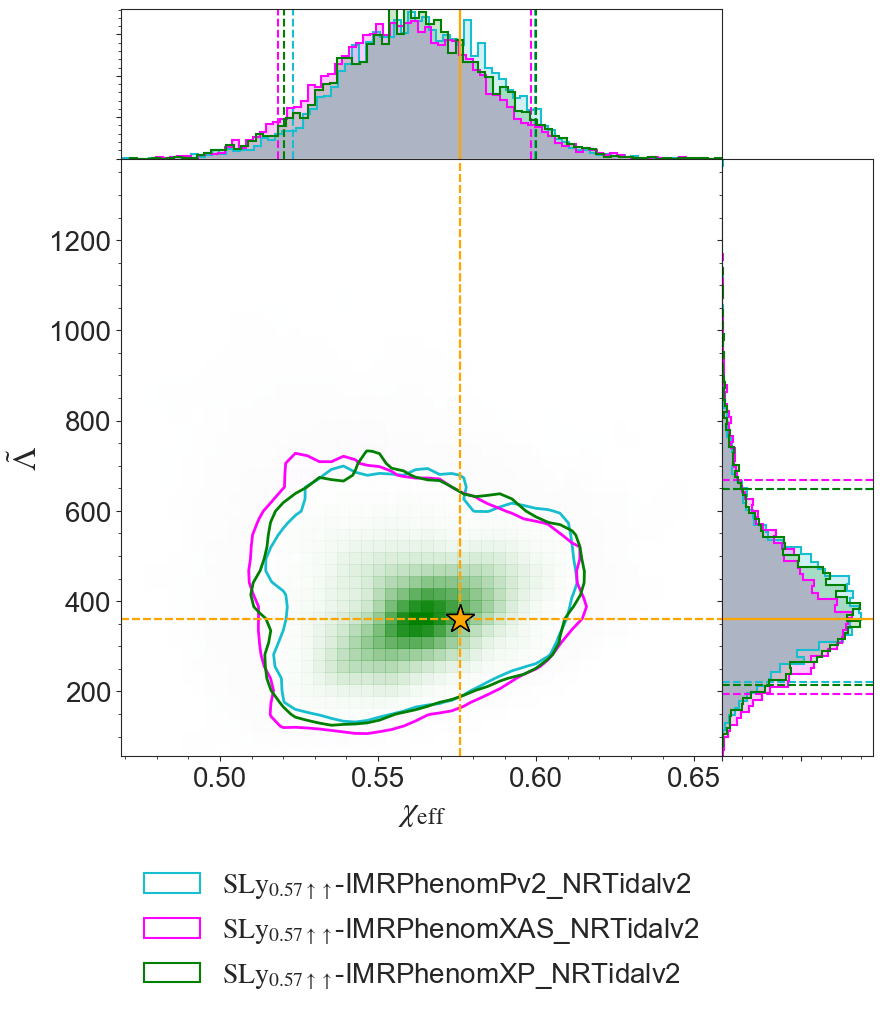}\\
\includegraphics[width=0.65\columnwidth]{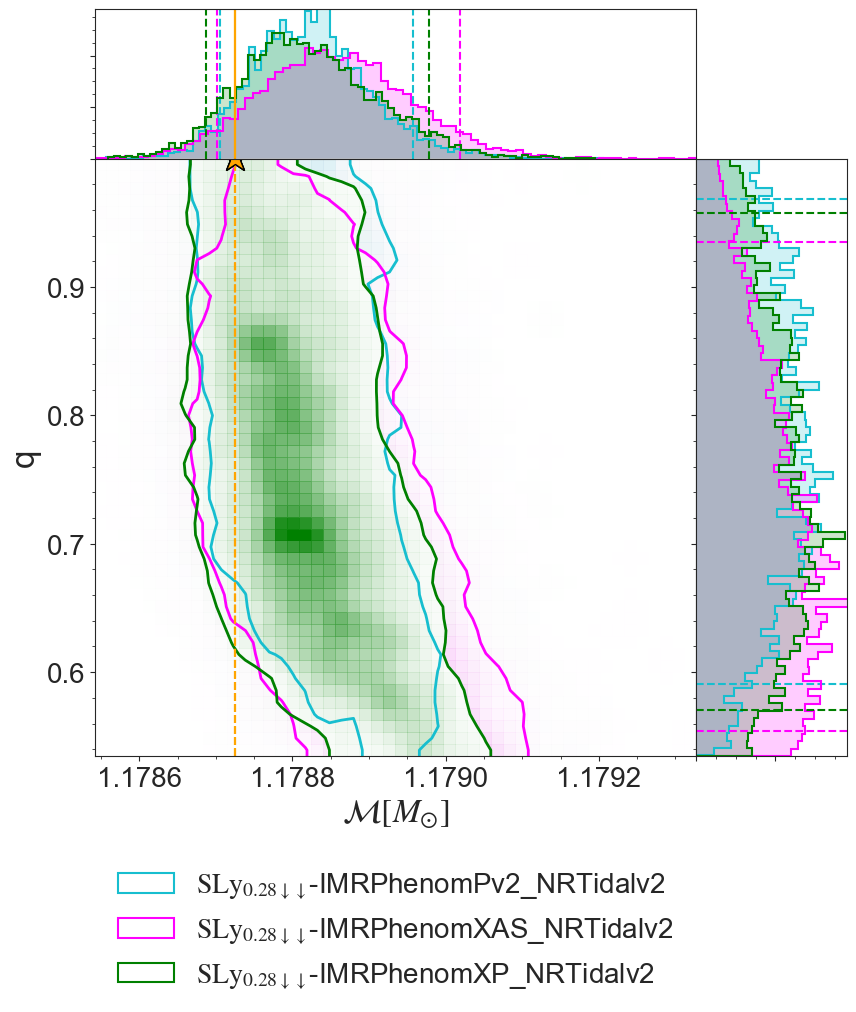}
\includegraphics[width=0.67\columnwidth]{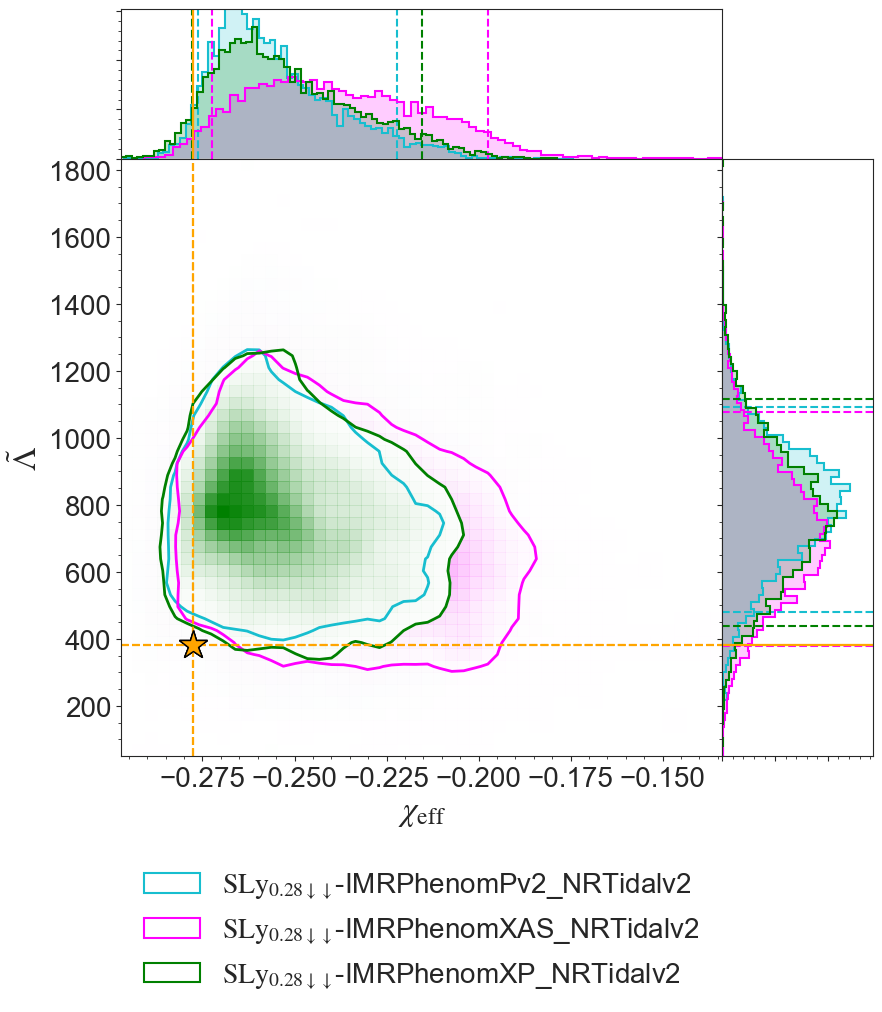}
\caption{1D and 2D posterior distributions for some key mass and spin parameters of two simulated signals created by hybridizing \textmd{SEOBNRv4T} and BAM waveforms (upper panels $\mathrm{SLy}_{0.57\uparrow\uparrow}$ and lower panels $\mathrm{SLy}_{0.28\downarrow\downarrow}$). Left panels show the 1D and 2D joint posterior distributions for the chirp mass $\mathcal{M}$ and mass ratio $q$; right panels show instead the distributions for the effective spin $\chi_{\rm{eff}}$ and the binary (mass-weighted) tidal deformability $\widetilde{\Lambda}$. The $q$ posteriors do not extend to lower values than are shown due to the restriction that the individual masses are $\geq 1\,\Msun$, as discussed in the text. Dashed (solid) lines mark the $90\%$ credible intervals of 1D (2D) posteriors and the true values of the parameters are marked by stars. Each plot compares the results obtained with three different waveform models: \textmd{PhenomPv2\_NRTidalv2} (cyan), \textmd{PhenomXAS\_NRTidalv2} (magenta), and \textmd{PhenomXP\_NRTidalv2} (green).
\label{fig:pe_bns_hybrids_inj}}
\end{center}
\end{figure*}

In this subsection, we focus on simulated signals emitted by BNSs with more significant spins. These signals were generated by hybridizing \textmd{SEOBNRv4T}~\cite{Hinderer:2016eia,Steinhoff:2016rfi} waveforms at low frequencies with BAM~\cite{Bruegmann:2006ulg,Thierfelder:2011yi,Dietrich:2015iva,Bernuzzi:2016pie} NR waveforms at high frequencies, and were first studied in Dudi \emph{et al.}~\cite{Dudi:2021wcf}. Here, we select two representative cases among the ones considered there, specifically $\mathrm{SLy}_{0.57\uparrow\uparrow}$ and $\mathrm{SLy}_{0.28\downarrow\downarrow}$  (see Table~I in Dudi \emph{et al.}). These cases have equal gravitational masses of $\sim1.35\,\Msun$ and equal aligned spins with the magnitudes and directions given in the case names. Both signals were injected at a fixed sky location corresponding to $\mathrm{RA}=5.5$~rad, $\mathrm{dec}=0.1$~rad, and geocentric GPS time $t_{0}=1239082262$~s, with an orbital inclination of $\iota=0$~rad, fixing the network signal-to-noise-ratio to $32.5$, in accordance with Dudi \emph{et al.} Based on these choices, the $\mathrm{SLy}_{0.57\uparrow\uparrow}$ ($\mathrm{SLy}_{0.28\downarrow\downarrow}$) signal was injected at a luminosity distance of $\sim 115.19$ ($112.87$)~Mpc. 
The hybrid waveforms were injected into a Hanford-Livingston-Virgo network in zero noise, and we employed the expected O4 noise curves for all detectors from~\cite{Aasi:2013wya}.
The simulated signals were analyzed over the frequency range $23$--$1024$~Hz and recovered with \textmd{PhenomXAS\_NRTidalv2}, \textmd{PhenomXP\_NRTidalv2}, and \textmd{PhenomPv2\_NRTidalv2}. As in our study of GW170817, we rely on the software \textmd{bilby}, sampling with \textmd{dynesty} and default sampler settings. Fig.~\ref{fig:pe_bns_hybrids_inj} shows the posterior distributions obtained for some of the mass, spin, and extrinsic parameters of the two configurations, with the upper (lower) row reporting the results for $\mathrm{SLy}_{0.57\uparrow\uparrow}$ ($\mathrm{SLy}_{0.28\downarrow\downarrow}$). We use \textmd{bilby}'s \textmd{UniformInComponentsChirpMass} and \textmd{UniformInComponentsMassRatio} priors, with prior bounds $\mathcal{M}\in[\mathcal{M}_{\rm{inj}}-0.02\Msun,\mathcal{M}_{\rm{inj}}+0.02\Msun]$ (where $\mathcal{M}_{\rm{inj}}$ is the injected value of chirp mass), $q\in[0.25,1]$. We place an additional constraint $1\,\Msun\leq m_{1,2}\leq 3\,\Msun$, whose bounds match those of the component mass prior employed in Dudi \emph{et al.}. This constraint leads to an additional restriction of the mass ratio in the posteriors we obtain.\footnote{We decided to place a lower bound of $1\,\Msun$ on the masses of neutron stars, since the minimum mass of a neutron star created by stellar core collapse is predicted to be slightly larger than $1\,\Msun$; see, e.g.,~\cite{Suwa:2018uni}. However, we acknowledge potential evidence for neutron stars with masses below $1\,\Msun$~\cite{Doroshenko:2022}.} For the spins, we place an upper bound of $0.9$ on the individual spin magnitudes, in line with Dudi \emph{et al.}; we use \textmd{bilby}'s AlignedSpin prior when running \textmd{PhenomXAS\_NRTidalv2}, whereas for precessing approximants we assume the spins are isotropically distributed as in the previous subsection. Finally, we impose priors on the individual tidal deformabilities $\Lambda_{1,2}\in[0,5000]$. 

The results of this study are shown in Fig.~\ref{fig:pe_bns_hybrids_inj} and are in line with those of the match study discussed in Sec.~\ref{subsub:aligned_matches}, i.e., current GW models are capable of correctly recovering the source parameters when the binary's spins are aligned with the orbital angular momentum, even for non-negligible spin magnitudes, as shown in the top panels of Fig.~\ref{fig:pe_bns_hybrids_inj}. On the other hand, for the configuration with anti-aligned spins, we observe a slight bias in the recovery of spin and tidal deformabilities, with the mass-weighted tidal deformability~\cite{Flanagan:2007ix,Favata:2013rwa} 
\begin{equation}
\tilde{\Lambda}=\frac{16}{3}\frac{(m_1+12 m_2)m_1^4\Lambda_1+(1\leftrightarrow2)}{(m_1+m_2)^5}
\end{equation} 
and $\chi_{\mathrm{eff}}$ being somewhat overestimated. In all cases, we can observe that $\chi_{\rm{eff}}=0$ is excluded, i.e., well outside the $90\%$ credible level. In the anti-aligned spin case, we also observe some differences between aligned-spin and precessing models, with \textmd{PhenomXAS\_NRTidalv2} returning broader posteriors for $\chi_\mathrm{eff}$ and $\mathcal{M}$. These results are consistent with the findings of Dudi \emph{et al.}, and in line with the study presented in Sec.~\ref{subsub:aligned_matches}, where we found that configurations with negative $\chi_\mathrm{eff}$ corresponded to the worst matches between \textmd{PhenomXAS\_NRTidalv2} and \textmd{SEOBNRv4T}.
We also observe an excellent agreement between the two precessing approximants considered here; there are a few noticeable differences with respect to the analysis Dudi \emph{et al.} (e.g., in the width of the $\tilde{\Lambda}$ posterior for the $\mathrm{SLy}_{0.57\uparrow\uparrow}$), which are likely due to different samplers being used. Possible differences in the parameter estimation setups also motivated our choice of reanalyzing the signals with \textmd{PhenomPv2\_NRTidalv2}, which was considered in~\cite{Dudi:2021wcf}.

\section{Conclusions}
\label{sec:conclusions}

In this paper, we have presented \phenxastidal and \phenxptidal, two new efficient waveform models for analyzing the GW signals emitted by BNS coalescences, coupling the NRTidalv2 tidal extension~\cite{Dietrich:2017aum, Dietrich:2018uni, Dietrich:2019kaq} to the state-of-the-art PhenomX waveform suite~\cite{Pratten:2020fqn,Garcia-Quiros:2020qlt,Pratten:2020ceb}.  \phenxastidal and \phenxptidal are publicly available through the algorithm library \textmd{LALSuite}~\cite{lalsuite} and they are suited to both aligned-spin and precessing systems, though they are restricted to the dominant $\ell=|m|=2$ coprecessing frame modes. Multibanding is activated by default, to maximize computational efficiency, making the default versions of these models the fastest currently available for BNS. For precessing systems, the user can choose among several options using different approximations to the PN spin-precession equations. By default, the current implementation relies on a approximation using precession averaging plus the leading correction, capturing double-spin effects. We have compared the evaluation times of different precession prescriptions for \phenxptidal, finding that they are all competitive when compared to other state-of-the-art models. In particular, we found that even a fully numerical solution of the orbit-averaged SpinTaylor equations (including the contribution of non-binary-black-hole spin-induced quadrupoles to the precession) can be embedded in the PhenomX twisting-up construction with a computational cost that, for total masses $\gtrsim 5\,\Msun$, is lower than that of \phenptidal, making it entirely affordable for future extensions to neutron star--black hole binaries, which might exhibit non-negligible precession more frequently than BNS.

A further speedup might be achieved by optimizing the numerical solution of the PN spin-precession equations \cite{Yu:2023lml}, the twisting-up algorithm \cite{Ramos-Buades:2023ehm}, or by means of reduced order models (which have already been created for precessing binary black hole waveforms~\cite{Gadre:2022sed}); it is also possible to include the non-binary-black-hole spin-induced quadrupole effects in the precession-averaged approximation~\cite{LaHaye:2022yxa}, and potentially also higher-order PN terms using a similar method. Applications to binary black hole systems are discussed in a companion paper \cite{Colleoni:2024knd}.

We have studied the impact of different precession prescriptions on the GW signal, finding that they differ the most for strongly inclined systems with non-negligible spins. If such an event was detected, we envisage that cross-checks with different precession prescriptions could be useful to correctly estimate the source properties (in particular, the matter effects on the spin-induced quadrupoles in the precession included in the SpinTaylor prescription will be important in some cases~\cite{Lyu:2023zxv}), though further PE studies, which we leave for future work, would be needed to fully validate these expectations. In our comparisons against other tidal models, we found that significant disagreement can be seen when the components are endowed with non-negligible spins, in line with previous findings. We established this through match studies as well as injections and recoveries of simulated highly spinning BNS signals. On the other hand, \phenxastidal and \phenxptidal appear in excellent agreement with other models when applied to low-spin binaries, such as GW170817.

The models presented here were used as the basis for the new IMRPhenomXAS\_NRTidalv3 and IMRPhenomXP\_NRTidalv3 models that provide a further calibration of the frequency-domain tidal phase including dynamical tides~\cite{Abac:2023ujg}, and could be extended further in several directions, through the phenomenological description of disruptive phenomena in neutron star--black hole binaries~\cite{Pannarale:2015jka,Thompson:2020nei}, the calibration of the precession to NR simulations (as is carried out in the \textmd{PhenomXO4a} binary black hole model~\cite{Thompson:2023ase,Hamilton:2021pkf}), the inclusion of the asymmetry of $\pm m$ modes in precessing cases (already included for the dominant $\ell=|m|=2$ modes in \textmd{PhenomXO4a}~\cite{Thompson:2023ase,Ghosh:2023mhc}), and the addition of subdominant modes \cite{Gamba:2023mww}.

\acknowledgements

We thank Sarp Akçay and Sebastiano Bernuzzi for providing useful comments on the manuscript. We thank Sarp Akçay, N.~V.~Krishnendu, and Shubhanshu Tiwari for their contributions during the LIGO-Virgo-KAGRA review of this work. We also thank Rossella Gamba for useful feedback about \textmd{TEOBResumS-GIOTTO} and for kindly implementing the code changes needed to make a direct comparison of frequency-domain precessing models.
M.C.\ acknowledges funding from the Spanish Agencia Estatal de Investigaci\'{o}n, grant IJC2019-041385. This work was supported by the Universitat de les Illes Balears (UIB); the Spanish Agencia Estatal de Investigaci\'{o}n, grants PID2022-138626NB-I00, PID2019-106416GB-I00, RED2022-134204-E, RED2022-134411-T, funded by MCIN/AEI/10.13039/501100011033; the MCIN with funding from the European Union NextGenerationEU/PRTR (PRTR-C17.I1); Comunitat Aut\`{o} noma de les Illes Balears through the Direcci\'{o} General de Recerca, Innovaci\'{o} I Transformaci\'{o} Digital with funds from the Tourist Stay Tax Law (PDR2020/11 - ITS2017-006), the Conselleria d’Economia, Hisenda i Innovaci\'{o} grant numbers SINCO2022/18146 and SINCO2022/6719, co-financed by the European Union and FEDER Operational Program 2021-2027 of the Balearic Islands; the “ERDF A way of making Europe.” N.K.J.-M.\ is supported by National Science Foundation (NSF) grant AST-2205920. G.P.\ gratefully acknowledges support from a Royal Society University Research Fellowship URF{\textbackslash}R1{\textbackslash}221500 and RF{\textbackslash}ERE{\textbackslash}221015, and from STFC grant ST/V005677/1. 
T.D.\ acknowledges support by the European Union (ERC, SMArt, 101076369). Views and opinions expressed are those of the authors only and do not necessarily reflect those of the European Union or the European Research Council. Neither the European Union nor the granting authority can be held responsible for them.
The authors are grateful for computational resources provided by Cardiff University and the LIGO Laboratory and supported by STFC grant ST/I006285/1 and NSF Grants PHY-0757058 and PHY-0823459. The authors thankfully acknowledge the computer resources at MareNostrum and Picasso, and
the technical support provided by Barcelona Supercomputing Center (BSC) through grants No. AECT-2023-2-0019 and AECT-2023-2-0032 from the Red Española Supercomputación (RES).
This research has made use of data obtained from the Gravitational Wave Open Science Center~\cite{LIGOScientific:2019lzm}, a service of LIGO Laboratory, the LIGO Scientific Collaboration, Virgo Collaboration and KAGRA. LIGO Laboratory and Advanced LIGO are funded by the United States NSF as well as the Science and Technology Facilities Council (STFC) of the United Kingdom, the Max-Planck-Society (MPS), and the State of Niedersachsen/Germany for support of the construction of Advanced LIGO and construction and operation of the GEO600 detector. Additional support for Advanced LIGO was provided by the Australian Research Council. Virgo is funded, through the European Gravitational Observatory (EGO), by the French Centre National de Recherche Scientifique (CNRS), the Italian Istituto Nazionale di Fisica Nucleare (INFN) and the Dutch Nikhef, with contributions by institutions from Belgium, Germany, Greece, Hungary, Ireland, Japan, Monaco, Poland, Portugal, Spain. KAGRA is supported by Ministry of Education, Culture, Sports, Science and Technology (MEXT), Japan Society for the Promotion of Science (JSPS) in Japan; National Research Foundation (NRF) and Ministry of Science and ICT (MSIT) in Korea; Academia Sinica (AS) and National Science and Technology Council (NSTC) in Taiwan.
This work made use of the software packages  \textmd{bilby}~\cite{Ashton:2018jfp}, \textmd{dynesty}~\cite{Speagle_2020}, \textmd{LALSuite}~\cite{lalsuite}, \textmd{matplotlib}~\cite{Hunter:2007ouj}, \textmd{numpy}~\cite{Harris:2020xlr}, \textmd{PyCBC}~\cite{PyCBC}, and \text{scipy}~\cite{Virtanen:2019joe}.

\appendix

\section{Settings for the model}
\label{app:settings}

\begin{table*}
\caption{New SpinTaylor options for computing the Euler angles. The versions with BBH SpinTaylor evolution are included for comparison to show the effect of tides and non-BH spin-induced quadrupoles on the precession. The differences between 31x and 32x are very slight for BNS signals.}
\begin{tabular}{cc}
\hline\hline
\emph{PhenomXPrecVersion} & Explanation\\
\hline
310 & SpinTaylor evolution with constant angles in the merger-ringdown phase.\\
311 & The same as 310 except with BBH SpinTaylor evolution.\\
320 (recommended) & SpinTaylor evolution with analytical continuation in the merger-ringdown phase.\\
321 & The same as 320 except with BBH SpinTaylor evolution.\\
\hline\hline
\end{tabular}
\label{tab:prec_settings}
\end{table*}

The new settings for precession (\textmd{PhenomXPrecVersion}) introduced here are given in Table~\ref{tab:prec_settings}. When the numerical integration of the SpinTaylor equations is stopped due to triggering one of its internal checks (that the evolution is self-consistent and producing sane results), the Euler angles are extended into merger-ringdown either setting them to a constant value (versions 31x), or by means of a $\mathrm{C}^1$ analytical continuation (versions 32x), with the latter being the recommended version. For example, the SpinTaylor evolution ends at $\sim 1840$~Hz for the single-spin example considered in Sec.~\ref{ssec:td_comp} (and one gets larger maximum frequencies for smaller tidal deformabilities).
One can see the other available settings in Table~III in~\cite{Pratten:2020ceb}. For the tests performed in this paper with SpinTaylor versions, we have set \textmd{PhenomXPFinalSpinMod} = 2. Using this option, the contribution of the in-plane spin components to the final spin is given by the norm of the total in-plane spin vector; the default final-spin option for \textmd{PhenomXP\_NRTidalv2} relies instead on precession-averaged quantities that are computed in the context of the MSA approximation and corresponds to \textmd{PhenomXPFinalSpinMod} = 3 (see Sec.~IV~D and Table~V of~\cite{Pratten:2020ceb} for further details). The other new parameter is the \textmd{PhenomXPSpinTaylorCoarseFactor} setting that determines the coarseness of the low-frequency grid used for the SpinTaylor evolution.

\section{Dependence of Euler angles on SpinTaylor integration settings}
\label{app:coarse_angles}
\begin{figure}[h]
\begin{center}
\includegraphics[width=\columnwidth]{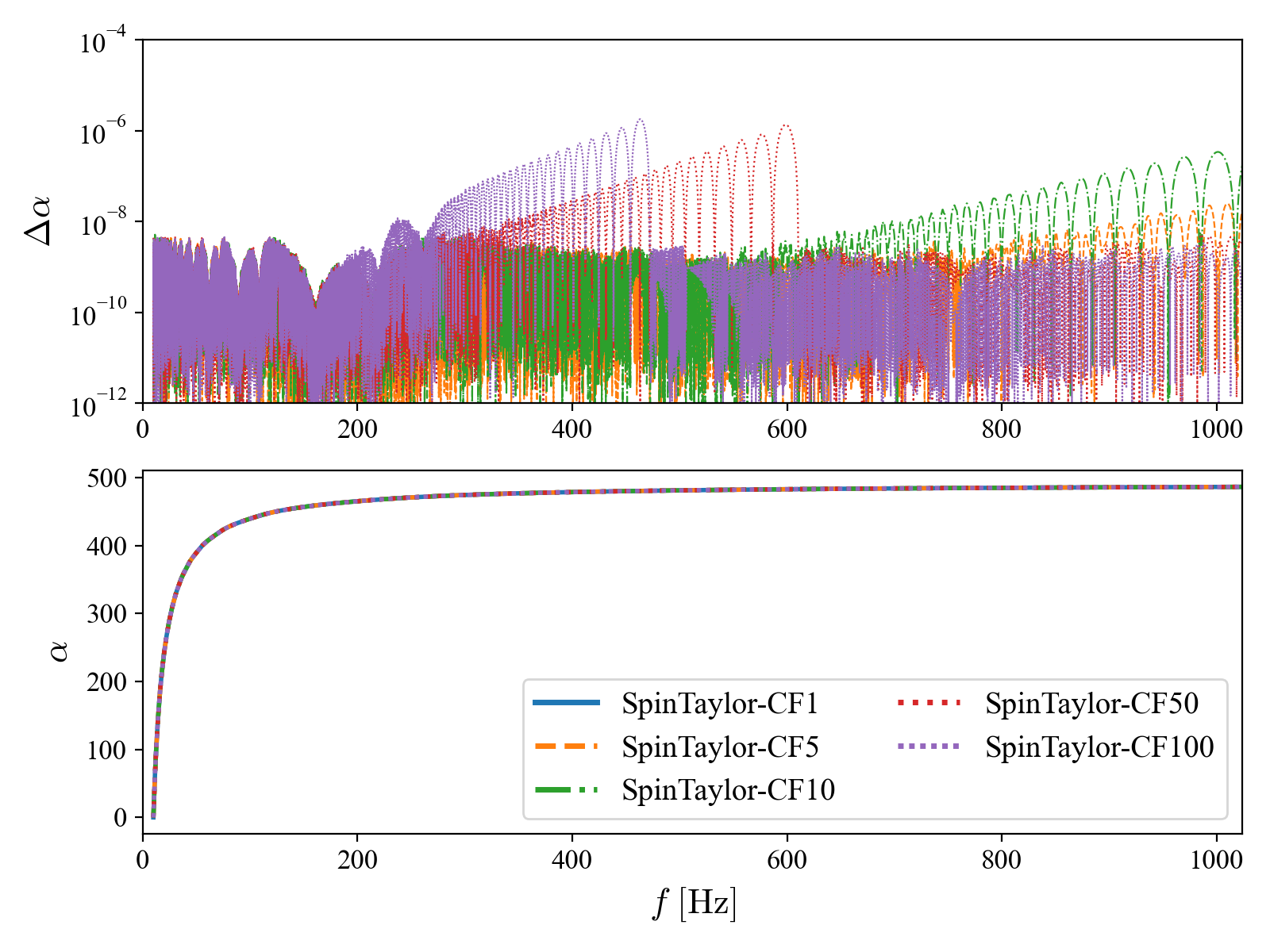}
\caption{The Euler angle $\alpha$ returned by the SpinTaylor model for an illustrative BNS system (lower panel; see Sec.~\ref{subsec:coarse_grid_integration} for further details) and the absolute errors with the respect to the uniform grid integration, denoted by CF$1$, for a variety of integration steps (upper panel), where CF$n$ corresponds to integration performed with a grid step $n$ times larger in the low frequency range. Note that in the upper panel we only show errors for CF$n$, where $n>1$, whereas the lower panel shows the tracks for all the cases in the legend. \label{fig:alpha_coarse}
}
\end{center}
\end{figure}

\begin{figure}[h]
\begin{center}
\includegraphics[width=\columnwidth]{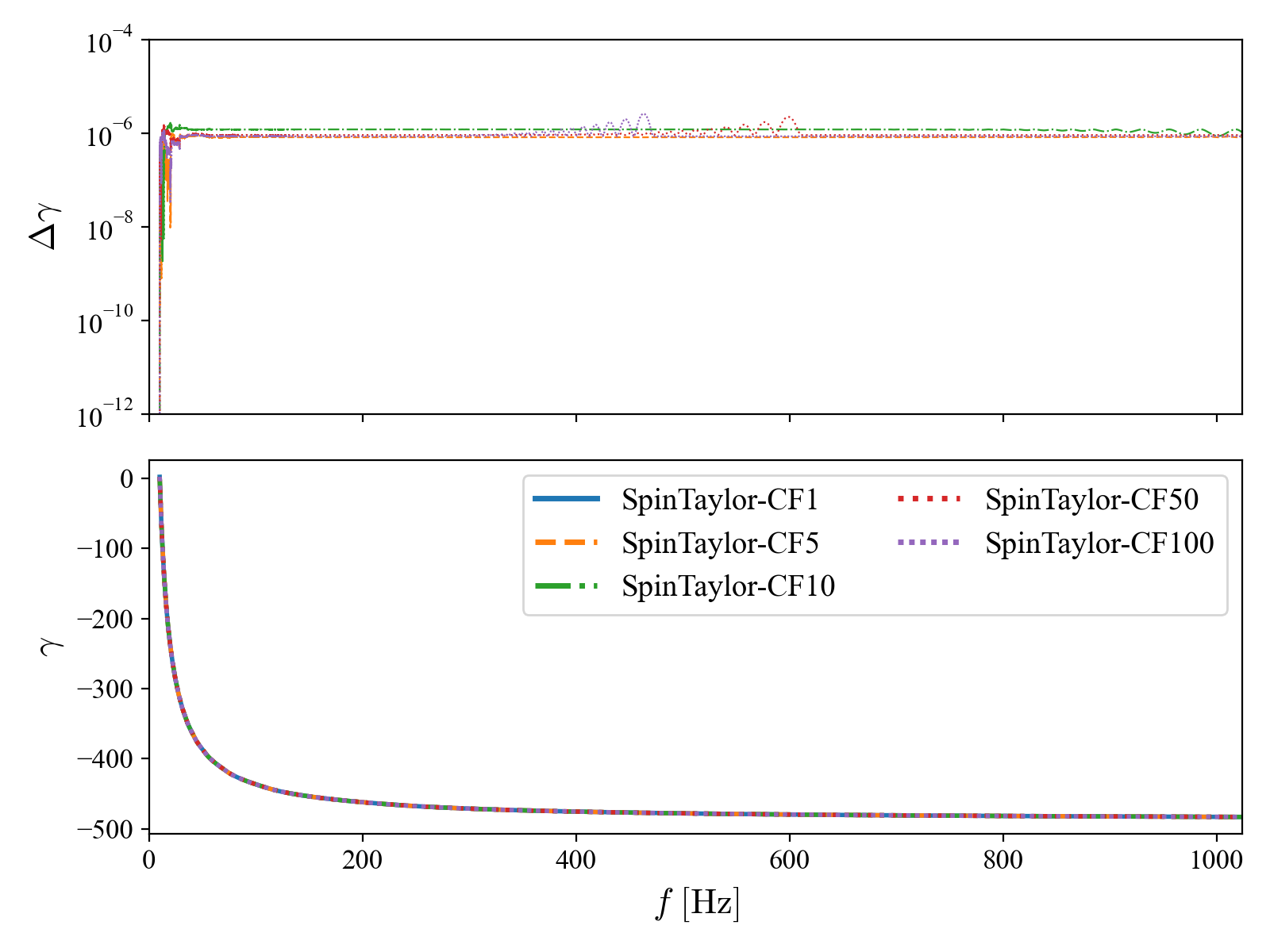}
\caption{The Euler angle $\gamma$ returned by the SpinTaylor model for an illustrative BNS system (lower panel; see Sec.~\ref{subsec:coarse_grid_integration} for further details) and the absolute errors with the respect to the uniform grid integration, denoted by CF$1$, for a variety of integration steps (upper panel), where CF$n$ corresponds to integration performed with a grid step $n$ times larger in the low frequency range. Note that in the upper panel we only show errors for CF$n$, where $n>1$, whereas the lower panel shows the tracks for all the cases in the legend. \label{fig:gamma_coarse}
}
\end{center}
\end{figure}

Figs.~\ref{fig:alpha_coarse} and~\ref{fig:gamma_coarse} show the absolute errors for the Euler angles $\alpha$ and $\gamma$ as a function of frequency for the same configuration discussed in Sec.~\ref{subsec:coarse_grid_integration}.

\bibliography{imrphenomxp_nrtidalv2}

\end{document}